\documentclass[%
reprint,
showpacs,preprintnumbers,
nofootinbib,
amsmath,amssymb,
aps,
prd,
floatfix,
]{revtex4-1}

\usepackage{graphicx}
\usepackage{dcolumn}
\usepackage{bm}

\usepackage{url}

\newcommand{\nua}[1]{\ensuremath{\rlap
           {\kern-2.5pt\ensuremath
           {\overset{\scriptscriptstyle(-)}{\phantom{\nu}}}}
           {\ensuremath{{\nu}_{#1}}}}}

\usepackage[ddmmyy,24hr]{datetime}

\usepackage{color}

\begin{document}

\preprint{\begin{tabular}{l}
\texttt{arXiv:1210.5715 [hep-ph]}
\end{tabular}}

\title{Update of Short-Baseline Electron Neutrino and Antineutrino Disappearance}

\author{C. Giunti}
\affiliation{INFN, Sezione di Torino, Via P. Giuria 1, I--10125 Torino, Italy}

\author{M. Laveder}
\affiliation{Dipartimento di Fisica e Astronomia ``G. Galilei'', Universit\`a di Padova,
and
INFN, Sezione di Padova,
Via F. Marzolo 8, I--35131 Padova, Italy}

\author{Y.F. Li}
\affiliation{Institute of High Energy Physics,
Chinese Academy of Sciences, Beijing 100049, China}

\author{Q.Y. Liu, H.W. Long}
\affiliation{Department of Modern Physics, University of Science and
Technology of China, Hefei, Anhui 230026, China}


\begin{abstract}
We present a complete update of the analysis of
$\nu_{e}$ and $\bar\nu_{e}$ disappearance experiments
in terms of neutrino oscillations
in the framework of 3+1 neutrino mixing,
taking into account
the Gallium anomaly,
the reactor anomaly,
solar neutrino data
and
$\nu_{e}\text{C}$ scattering data.
We discuss the implications of a recent
${}^{71}\text{Ga}({}^{3}\text{He},{}^{3}\text{H}){}^{71}\text{Ge}$
measurement
which give information on the neutrino cross section in Gallium experiments.
We discuss the solar bound
on active-sterile mixing
and present our numerical results.
We discuss the
connection between the results of the fit of neutrino oscillation data
and the heavy neutrino mass effects in
$\beta$-decay experiments
(considering new Mainz data)
and
neutrinoless double-$\beta$ decay experiments
(considering the recent EXO results).
\end{abstract}


\maketitle

\section{Introduction}
\label{Introduction}

In recent years several short-baseline neutrino oscillation experiments have found anomalies which
may require an extension of the standard three-neutrino mixing framework
which describes the neutrino oscillations observed in
solar, atmospheric and long-baseline experiments
(see \cite{Giunti:2007ry,Bilenky:2010zza,Xing:2011zza}).
In this paper we consider the Gallium anomaly
\cite{Laveder:2007zz,hep-ph/0610352,Giunti:2010zu}
and
the reactor anomaly
\cite{Mueller:2011nm,Huber:2011wv,Mention:2011rk},
which indicate that electron neutrino and antineutrinos
may disappear at short distances\footnote{
The inclusion in the analysis of the more controversial
LSND
\cite{hep-ex/0104049}
and
MiniBooNE
\cite{1207.4809}
$\nua{\mu}\to\nua{e}$
anomalies
will be discussed elsewhere
\cite{Giunti-Laveder-Li-Long-12}.}.
Such disappearance may be explained by the presence of at least
one massive neutrino at the eV scale,
which drives short-baseline neutrino oscillations
generated by a squared-mass difference which is much larger than
the squared-mass difference operating in
the solar, atmospheric and long-baseline neutrino oscillation experiments.
We consider 3+1 neutrino mixing,
which is the minimal extension of three-neutrino mixing which can explain the
Gallium and reactor anomalies.
Since from the LEP measurement of the
invisible width of the $Z$ boson
\cite{hep-ex/0509008}
we know that there are only three light active flavor neutrinos
the additional neutrino in the 3+1 framework is sterile.

In this paper we discuss the implications of
the recent
${}^{71}\text{Ga}({}^{3}\text{He},{}^{3}\text{H}){}^{71}\text{Ge}$
measurement
in Ref.~\cite{Frekers:2011zz},
which give information on the neutrino cross section in Gallium experiments.
We take also into account the most updated
calculation of the reactor neutrino fluxes presented in the recent White Paper on
light sterile neutrinos
\cite{Abazajian:2012ys}.
We present also a detailed discussion of the
connection between the results of the fit of neutrino oscillation data
and the results of
$\beta$-decay experiments
(considering the Mainz data presented very recently in Ref.~\cite{1210.4194})
and
neutrinoless double-$\beta$ decay experiments
(considering the recent EXO bound in Ref.~\cite{Auger:2012ar}
and the controversial positive result in Ref.~\cite{KlapdorKleingrothaus:2006ff}).

We consider 3+1 neutrino mixing as an extension of standard
three-neutrino mixing.
The mixing of the three active flavor neutrino fields
$\nu_{e}$,
$\nu_{\mu}$,
$\nu_{\tau}$
and one sterile neutrino field $\nu_{s}$
is given by
\begin{equation}
\nu_{\alpha}
=
\sum_{k=1}^{4} U_{\alpha k} \nu_{k}
\,,
\label{mixing}
\end{equation}
where
$U$ is the unitary $4\times4$ mixing matrix
($U^{\dagger}=U^{-1}$)
and
each of the four
$\nu_{k}$'s is a massive neutrino field with mass $m_{k}$.
We consider the squared-mass hierarchy
\begin{equation}
\Delta{m}^{2}_{21}
\ll
\Delta{m}^{2}_{31}
\ll
\Delta{m}^{2}_{41}
\,,
\label{dm2hierarchy}
\end{equation}
with
$\Delta{m}^{2}_{kj}\equiv m_{k}^2 - m_{j}^2$,
such that
$\Delta{m}^{2}_{21}$
generates the very-long-baseline oscillations observed
in solar neutrino experiments and in the KamLAND reactor antineutrino experiment,
$\Delta{m}^{2}_{31}$
generates the long-baseline oscillations observed
in atmospheric neutrino experiments and in long-baseline accelerator and reactor neutrino and antineutrino experiments,
and
$\Delta{m}^{2}_{41}$
generates short-baseline oscillations.

The effective survival probability
at a distance $L$ of electron neutrinos and antineutrinos with energy $E$
in short-baseline neutrino oscillation experiments
is given by
(see Refs.~\cite{hep-ph/9812360,hep-ph/0405172,hep-ph/0606054,GonzalezGarcia:2007ib})
\begin{equation}
P_{\nua{e}\to\nua{e}}^{\text{SBL}}
=
1
-
\sin^{2}2\vartheta_{ee}
\sin^{2}\left( \frac{\Delta{m}^2_{41} L}{4E} \right)
\,,
\label{survi}
\end{equation}
with the transition amplitude
\begin{equation}
\sin^{2}2\vartheta_{ee}
=
4 |U_{e4}|^2 \left( 1 - |U_{e4}|^2 \right)
\,.
\label{survisin}
\end{equation}

The plan of the paper is as follows.
In Section~\ref{Gallium Anomaly}
we discuss in detail the Gallium $\nu_{e}$ anomaly
\cite{Laveder:2007zz,hep-ph/0610352,Giunti:2010zu}
and the implications of the important recent
${}^{71}\text{Ga}({}^{3}\text{He},{}^{3}\text{H}){}^{71}\text{Ge}$
measurement
in Ref.~\cite{Frekers:2011zz}.
In Section~\ref{Fit of Gallium and reactor data}
we present the results of the combined analysis of Gallium data
with reactor $\bar\nu_{e}$ data,
taking into account the reactor $\bar\nu_{e}$ anomaly
\cite{Mueller:2011nm,Huber:2011wv,Mention:2011rk,Abazajian:2012ys}.
In Section~\ref{Solar neutrino constrait} we discuss the
solar neutrino constraint on short-baseline $\nu_{e}$ disappearance
\cite{Giunti:2009xz,Palazzo:2011rj,Palazzo:2012yf,Palazzo-NOW2012}.
In Section~\ref{Global fit}
we present the results of the global fit of
$\nu_{e}$ and $\bar\nu_{e}$ disappearance data,
which includes also
$\nu_{e} + {}^{12}\text{C} \to {}^{12}\text{N}_{\text{g.s.}} + e^{-}$
scattering data \cite{1106.5552,Giunti:2011cp}.
We confront these results with the bounds on the heavy neutrino mass
given by the data of
$\beta$-decay experiments \cite{Galeazzi:2001py,1210.4194}
and
neutrinoless double-$\beta$ decay experiments
\cite{Auger:2012ar,KlapdorKleingrothaus:2006ff}.
Finally,
in Section~\ref{Conclusions} we draw our conclusions.

\begin{table}[b]
\caption{
\label{tab:src}
Energy ($E$) and branching ratio ($\text{B.R.}$)
of the neutrino lines produced in the electron-capture decay of
${}^{51}\text{Cr}$ and ${}^{37}\text{Ar}$.
}
\begin{ruledtabular}
\begin{tabular}{l|cccc|cc}
&
\multicolumn{4}{c|}{${}^{51}\text{Cr}$}
&
\multicolumn{2}{c}{${}^{37}\text{Ar}$}
\\
$E\,[\text{keV}]$	& $  747 $ & $  752 $ & $  427 $ & $  432 $ & $  811$ & $  813$
\\
$\text{B.R.}$		& $0.8163$ & $0.0849$ & $0.0895$ & $0.0093$ & $0.902$ & $0.098$
\end{tabular}
\end{ruledtabular}
\end{table}

\begin{table}[b]
\caption{
\label{tab:rat}
Ratios of measured and expected ${}^{71}\text{Ge}$ event rates
in the four radioactive source experiments.
$\text{G1}$ and $\text{G2}$ denote the two GALLEX experiments with ${}^{51}\text{Cr}$ sources
\protect\cite{Anselmann:1995ar,Hampel:1998fc,1001.2731},
$\text{S1}$ denotes the SAGE experiment with a ${}^{51}\text{Cr}$ source,
and
$\text{S2}$ denotes the SAGE experiment with a ${}^{37}\text{Ar}$ source
\protect\cite{Abdurashitov:1996dp,hep-ph/9803418,nucl-ex/0512041,0901.2200}.
AVE denotes the weighted average.
}
\begin{ruledtabular}
\begin{tabular}{lccccc}
		& G1																		  & G2																		  & S1																		  & S2																		  & AVE                                                                                                                                          
\\
$R_{\text{B}}$	& $0.95 {}^{+0.11}_{-0.11}															$ & $0.81 {}^{+0.10}_{-0.11}															$ & $0.95 {}^{+0.12}_{-0.12}															$ & $0.79 {}^{+0.08}_{-0.08}															$ & $0.86{}^{+0.05}_{-0.05}	$
\\
$R_{\text{HK}}$	& $0.85{}^{+0.12}_{-0.12}	$ & $0.71{}^{+0.11}_{-0.11}$ & $0.84{}^{+0.13}_{-0.12}$ & $0.71{}^{+0.09}_{-0.09}$ & $0.77{}^{+0.08}_{-0.08}$
\\
$R_{\text{FF}}$	& $0.93{}^{+0.11}_{-0.11}	$ & $0.79{}^{+0.10}_{-0.11}$ & $0.93{}^{+0.11}_{-0.12}$ & $0.77{}^{+0.09}_{-0.07}$ & $0.84{}^{+0.05}_{-0.05}$
\\
$R_{\text{HF}}$	& $0.83{}^{+0.13}_{-0.11}	$ & $0.71{}^{+0.11}_{-0.11}$ & $0.83{}^{+0.13}_{-0.12}$ & $0.69{}^{+0.10}_{-0.09}$ & $0.75{}^{+0.09}_{-0.07}$
\end{tabular}
\end{ruledtabular}
\end{table}

\section{Gallium Anomaly}
\label{Gallium Anomaly}

The GALLEX
\cite{Anselmann:1995ar,Hampel:1998fc,1001.2731}
and
SAGE
\cite{Abdurashitov:1996dp,hep-ph/9803418,nucl-ex/0512041,0901.2200}
Gallium solar neutrino experiments have been tested with
intense artificial ${}^{51}\text{Cr}$ and ${}^{37}\text{Ar}$ radioactive sources
which produce electron neutrinos through electron capture
with the energies and branching ratios given in
Tab.~\ref{tab:src}.
In each of these experiments
the source was placed near the center of the approximately cylindrical detector
and electron neutrinos have been detected with the solar neutrino detection reaction
\begin{equation}
\nu_{e} + {}^{71}\text{Ga} \to {}^{71}\text{Ge} + e^{-}
\,.
\label{f31}
\end{equation}
The average neutrino travelling distances are
$\langle L \rangle_{\text{GALLEX}} = 1.9 \, \text{m}$
and
$\langle L \rangle_{\text{SAGE}} = 0.6 \, \text{m}$.
The first line in Tab.~\ref{tab:rat}
shows the ratios $R_{\text{B}}$ of measured and expected ${}^{71}\text{Ge}$ event rates
reported by the experimental collaborations.
The index B indicates that the expected event rates have been calculated using the
Bahcall cross sections \cite{Bahcall:1997eg}
\begin{align}
\sigma_{\text{B}}({}^{51}\text{Cr})
=
\null & \null
58.1 \times 10^{-46} \, \text{cm}^2
\,,
\label{BCr}
\\
\sigma_{\text{B}}({}^{37}\text{Ar})
=
\null & \null
70.0 \times 10^{-46} \, \text{cm}^2
\,,
\label{BAr}
\end{align}
without considering their uncertainties.
One can see that the values of
$R_{\text{B}}^{\text{G1}}$ and $R_{\text{B}}^{\text{S1}}$
indicate a compatibility between the measured and expected event rates,
whereas the values of
$R_{\text{B}}^{\text{G2}}$ and $R_{\text{B}}^{\text{S2}}$
are significantly smaller than one,
indicating a disappearance of electron neutrinos.
The weighted average in Tab.~\ref{tab:rat} gives a
$2.7\sigma$
anomaly.

Since the values of the cross sections of
${}^{51}\text{Cr}$ and ${}^{37}\text{Ar}$
electron neutrinos and their uncertainties
are crucial for the interpretation of the Gallium data
as indication of short-baseline $\nu_{e}$ disappearance,
in the following we discuss in detail the problem of the determination of the cross sections
and their uncertainties,
taking into account Refs.~\cite{Krofcheck:1985fg,Krofcheck-PhD-1987,Bahcall:1997eg,Haxton:1998uc}
and the important recent measurement in Ref.~\cite{Frekers:2011zz}.

\begin{figure}[t]
\begin{center}
\includegraphics*[width=0.8\linewidth]{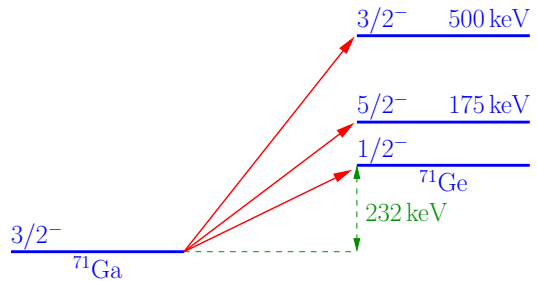}
\end{center}
\caption{ \label{fig:GaGe}
${}^{71}\text{Ga}\to{}^{71}\text{Ge}$
transitions induced by
${}^{51}\text{Cr}$
and
${}^{37}\text{Ar}$
electron neutrinos.
}
\end{figure}

The cross sections of the interaction process (\ref{f31})
for neutrinos produced by
${}^{51}\text{Cr}$ and ${}^{37}\text{Ar}$
sources are given by
\begin{equation}
\sigma
=
\sigma_{\text{gs}}
\left(
1
+
\xi_{175}
\frac{\text{BGT}_{175}}{\text{BGT}_{\text{gs}}}
+
\xi_{500}
\frac{\text{BGT}_{500}}{\text{BGT}_{\text{gs}}}
\right)
\,,
\label{cs01}
\end{equation}
where
$\sigma_{\text{gs}}$
is the cross sections of the transitions
from the ground state of ${}^{71}\text{Ga}$ to the ground state of ${}^{71}\text{Ge}$,
$\text{BGT}_{\text{gs}}$
is the corresponding Gamow-Teller strength,
and
$\text{BGT}_{175}$
and
$\text{BGT}_{500}$
are the Gamow-Teller strengths of the transitions
from the ground state of ${}^{71}\text{Ga}$ to the two excited states of ${}^{71}\text{Ge}$
at about 175 keV and 500 keV (see Fig.~\ref{fig:GaGe}).
The coefficients of
$\text{BGT}_{175}/\text{BGT}_{\text{gs}}$
and
$\text{BGT}_{500}/\text{BGT}_{\text{gs}}$
are determined by phase space:
$\xi_{175}({}^{51}\text{Cr}) = 0.669$,
$\xi_{500}({}^{51}\text{Cr}) = 0.220$,
$\xi_{175}({}^{37}\text{Ar}) = 0.695$,
$\xi_{500}({}^{37}\text{Ar}) = 0.263$
\cite{Bahcall:1997eg}.

The cross sections of the transitions
from the ground state of ${}^{71}\text{Ga}$ to the ground state of ${}^{71}\text{Ge}$
have been calculated accurately by Bahcall
\cite{Bahcall:1997eg}:
\begin{align}
\sigma_{\text{gs}}({}^{51}\text{Cr})
=
55.3 \times 10^{-46} \, \text{cm}^2
\,,
\label{f23}
\\
\sigma_{\text{gs}}({}^{37}\text{Ar})
=
66.2 \times 10^{-46} \, \text{cm}^2
\,.
\label{f24}
\end{align}
These cross sections are proportional to the characteristic neutrino absorption cross section
\cite{Bahcall:1989ks,Bahcall:1997eg}
\begin{align}
\sigma_{0}
=
\null & \null
2 \alpha Z_{\text{Ge}} m_{e}^2 G_{\text{F}}^2 |V_{ud}|^2 g_{A}^2 \text{BGT}_{\text{gs}}
\nonumber
\\
=
\null & \null
Z_{\text{Ge}} \text{BGT}_{\text{gs}}
\,
(3.091 \pm 0.012) \times 10^{-46} \, \text{cm}^2
\,,
\label{f14}
\end{align}
where
$\alpha$ is the fine-structure constant,
$Z_{\text{Ge}}=32$ is the atomic number of the final nucleus,
$m_{e}$ is the electron mass,
$G_{\text{F}}$ is the Fermi constant,
$V_{ud}$ is the $ud$ element of the quark mixing matrix $V$, and
$g_{A}$ is the axial coupling constant.
The numerical value of the coefficient in the last line of Eq.~(\ref{f14})
has been obtained with the values of these quantities given in
the last Review of Particle Physics \cite{PDG-2012}.
From the value
\begin{equation}
\sigma_{0}
=
(8.611 \pm 0.011) \times 10^{-46} \, \text{cm}^2
\,,
\label{f21a}
\end{equation}
calculated by Bahcall
\cite{Bahcall:1997eg}
using the accurate measurement
\cite{Hampel:1985zz}
\begin{equation}
T_{1/2}({}^{71}\text{Ge}) = 11.43 \pm 0.03 \, \text{d}
\label{f03}
\end{equation}
of the lifetime of ${}^{71}\text{Ge}$
(which decays through the electron-capture process
$
e^{-} + {}^{71}_{32}\text{Ge} \to {}^{71}_{31}\text{Ga} + \nu_{e}
$,
which is the inverse of the $\nu_{e}$ detection process (\ref{f31})),
we obtain
\begin{equation}
\text{BGT}_{\text{gs}}
=
0.0871 \pm 0.0004
\,.
\label{f32}
\end{equation}
This value agrees with that given in Ref.~\cite{Haxton:1998uc},
but it is different from that recommended in Ref.~\cite{Frekers:2011zz}.
Hence,
we checked it using the relation
\begin{align}
\text{BGT}_{\text{gs}}
=
\null & \null
\frac{\left[2J_{\text{Ge}}+1\right]}{\left[2J_{\text{Ga}}+1\right]}
\,
\frac{2 \pi^3 \ln2}{G_{\text{F}}^2 |V_{ud}|^2 m_{e}^5 g_{A}^2 ft_{1/2}({}^{71}\text{Ge})}
\nonumber
\\
=
\null & \null
\frac{6289 \pm 3 \, \text{s}}{2 g_{A}^2 ft_{1/2}({}^{71}\text{Ge})}
\,,
\label{bgt-ft}
\end{align}
with
$J_{\text{Ge}}=1/2$
and
$J_{\text{Ga}}=3/2$,
and the value
\begin{equation}
\log ft_{1/2}({}^{71}\text{Ge})
=
4.3493 \pm 0.0015
\,,
\label{f04}
\end{equation}
obtained with
the LOGFT calculator
\cite{LOGFT-2012}
of the National Nuclear Data Center
using the lifetime (\ref{f03}).
The result,
\begin{equation}
\text{BGT}_{\text{gs}}
=
0.0872 \pm 0.0005
\,,
\label{f06}
\end{equation}
is in agreement with the value (\ref{f32}),
which will be used in the following.

\begin{table*}[t]
\caption{
\label{tab:bgt}
Values of the Gamow-Teller strengths of the transitions
from the ground state of ${}^{71}\text{Ga}$ to the two excited states of ${}^{71}\text{Ge}$
at 175 keV and 500 keV
and their relative values with respect to the
Gamow-Teller strength of the transitions to the ground state of ${}^{71}\text{Ge}$,
given in Eq.~(\ref{f32}).
}
\begin{ruledtabular}
\begin{tabular}{lccccc}
Reference
&
Method
&
$\text{BGT}_{175}$
&
$\dfrac{ \text{BGT}_{175} }{ \text{BGT}_{\text{gs}} }$
&
$\text{BGT}_{500}$
&
$\dfrac{ \text{BGT}_{500} }{ \text{BGT}_{\text{gs}} }$
\\
Krofcheck et al.
\protect\cite{Krofcheck:1985fg,Krofcheck-PhD-1987}
&
${}^{71}\text{Ga} (p,n) {}^{71}\text{Ge}$
&
$< 0.005$
&
$< 0.057$
&
$
0.011
\pm
0.002
$
&
$
0.126
\pm
0.023
$
\\
Haxton
\protect\cite{Haxton:1998uc}
&
Shell Model
&
$
0.017
\pm
0.015
$
&
$
0.19
\pm
0.18
$
&
&
\\
Frekers et al.
\protect\cite{Frekers:2011zz}
&
${}^{71}\text{Ga} ({}^{3}\text{He},{}^{3}\text{H}) {}^{71}\text{Ge}$
&
$
0.0034
\pm
0.0026
$
&
$
0.039
\pm
0.030
$
&
$
0.0176
\pm
0.0014
$
&
$
0.202
\pm
0.016
$
\end{tabular}
\end{ruledtabular}
\end{table*}

The Gamow-Teller strengths
$\text{BGT}_{175}$
and
$\text{BGT}_{500}$
have been measured in 1985 in the
$(p,n)$
experiment of Krofcheck et al. \cite{Krofcheck:1985fg,Krofcheck-PhD-1987}
and recently, in 2011,
in the
$({}^{3}\text{He},{}^{3}\text{H})$
experiment of Frekers et al.
\cite{Frekers:2011zz}.
The results are listed in Tab.~\ref{tab:bgt}
together with the 1998 shell-model calculation of
$\text{BGT}_{175}$
of Haxton
\cite{Haxton:1998uc}.

The Bahcall cross sections (\ref{BCr}) and (\ref{BAr})
have been obtained using for
$\text{BGT}_{500}$
the Krofcheck et al.
measurement
and
for
$\text{BGT}_{175}$
half of the Krofcheck et al.
upper limit
\cite{Bahcall:1997eg}.

\begin{table}[b]
\caption{
\label{tab:cs}
Gallium cross section (in units of $10^{-46} \, \text{cm}^2$) and its ratio with the corresponding Bahcall cross section (Eqs.~(\ref{BCr}) and (\ref{BAr}))
for ${}^{51}\text{Cr}$ and ${}^{37}\text{Ar}$ neutrinos
in the three cases discussed in the text.
}
\begin{ruledtabular}
\begin{tabular}{l|cc|cc}
&
\multicolumn{2}{c|}{${}^{51}\text{Cr}$}
&
\multicolumn{2}{c}{${}^{37}\text{Ar}$}
\\
   & $\sigma										$ & $\sigma/\sigma_{\text{B}}								$ & $\sigma										$ & $\sigma/\sigma_{\text{B}}								$
\\
HK & $63.9\pm6.5	$ & $1.10\pm0.11	$ & $77.2\pm8.1	$ & $1.10\pm0.12	$
\\
FF & $59.2\pm1.1	$ & $1.02\pm0.02	$ & $71.5\pm1.4	$ & $1.02\pm0.02	$
\\
HF & $64.9\pm6.5	$ & $1.12\pm0.11	$ & $78.5\pm8.1	$ & $1.12\pm0.12	$
\end{tabular}
\end{ruledtabular}
\end{table}

In previous publications
\cite{Giunti:2010zu,1008.4750,1107.1452,1109.4033,Giunti:2011cp,1207.6515}
we used the Haxton shell-model value of
$\text{BGT}_{175}$
and the $(p,n)$ measured value of
$\text{BGT}_{500}$.
Although the uncertainties of the Haxton shell-model value of
$\text{BGT}_{175}$
are so large that $\text{BGT}_{175}$ may be negligibly small,
the central value is much larger than the upper limit obtained in the
$(p,n)$ experiment.
According to Ref.~\cite{Haxton:1998uc},
this is due to a suppression of the
$(p,n)$ value caused by a destructive interference between the spin
($\Delta{J}=1$, $\Delta{L}=0$)
matrix element and an additional
spin-tensor
($\Delta{J}=1$, $\Delta{L}=2$)
matrix element
which operates only in $(p,n)$ transitions.
We do not know if the same suppression is operating also in
$({}^{3}\text{He},{}^{3}\text{H})$,
which could be the explanation of the smallness of the value of
$\text{BGT}_{175}$
measured by Frekers et al.,
which is compatible with the $(p,n)$ upper bound.
Moreover,
the value of $\text{BGT}_{500}$
measured by Frekers et al.
has a
$2.7\sigma$
discrepancy with that measured by Krofcheck et al.
Since we cannot solve these problems,
we consider the following three approaches
which give the cross sections in Tab.~\ref{tab:cs}
and the ratios of measured and expected ${}^{71}\text{Ge}$ event rates in Tab.~\ref{tab:rat}:

\begin{description}

\item[HK]
Haxton $\text{BGT}_{175}$ value
and
Krofcheck et al. $\text{BGT}_{500}$ value.
This is our old approach adopted in previous publications
\cite{Giunti:2010zu,1008.4750,1107.1452,1109.4033,Giunti:2011cp,1207.6515}.
The cross sections are significantly larger than the Bahcall cross sections in Eqs.~(\ref{BCr}) and (\ref{BAr}),
albeit with large uncertainties which make them compatible at the $1\sigma$ level.

\item[FF]
Frekers et al. values of both
$\text{BGT}_{175}$
and
$\text{BGT}_{500}$.
This is a new approach which is motivated by the new $({}^{3}\text{He},{}^{3}\text{H})$
measurements
\cite{Frekers:2011zz}.
The cross sections are only slightly larger than the Bahcall cross sections in Eqs.~(\ref{BCr}) and (\ref{BAr}),
mainly because of the larger $\text{BGT}_{500}$.

\item[HF]
Haxton $\text{BGT}_{175}$ value
and
Frekers et al. $\text{BGT}_{500}$ value.
This is a new approach which is motivated by the possibility that
the $\text{BGT}_{175}$ measured by Frekers et al.
suffers of destructive interference between the spin and spin-tensor
matrix element
and its value is different from the $\text{BGT}_{175}$ in Gallium neutrino detection,
as discussed by Haxton for the $(p,n)$ experiment \cite{Haxton:1998uc}.
This approach gives the largest cross sections,
which however are still compatible with the Bahcall cross sections
at the $1\sigma$ level.

\end{description}

From the weighted averages of measured and expected ${}^{71}\text{Ge}$ event rates in Tab.~\ref{tab:rat}
it follows that the statistical significance of the Gallium anomaly in the three cases is, respectively,
about
$3.0\sigma$,
$2.9\sigma$ and
$3.1\sigma$.
Hence,
the new $({}^{3}\text{He},{}^{3}\text{H})$
cross section measurement of Frekers et al.
\cite{Frekers:2011zz}
confirm that
there is a Gallium anomaly at a level of about $3\sigma$
\cite{Giunti:2010zu},
which indicates a short-baseline disappearance of $\nu_{e}$
which can be explained by neutrino oscillations.

\begin{figure}[t]
\begin{center}
\includegraphics*[width=\linewidth]{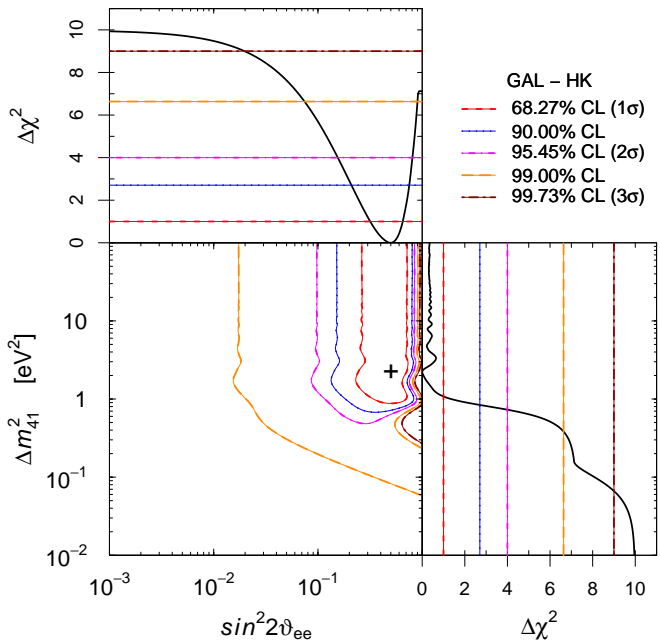}
\end{center}
\caption{ \label{fig:fit-ghk}
Allowed regions in the
$\sin^{2}2\vartheta_{ee}$--$\Delta{m}^{2}_{41}$ plane
and
marginal $\Delta\chi^{2}$'s
for
$\sin^{2}2\vartheta_{ee}$ and $\Delta{m}^{2}_{41}$
obtained from the
combined fit of the results of
the Gallium radioactive source experiments
in the HK case (see the text).
The best-fit point corresponding to $\chi^2_{\text{min}}$ is indicated by a cross.
}
\end{figure}

\begin{figure}[t]
\begin{center}
\includegraphics*[width=\linewidth]{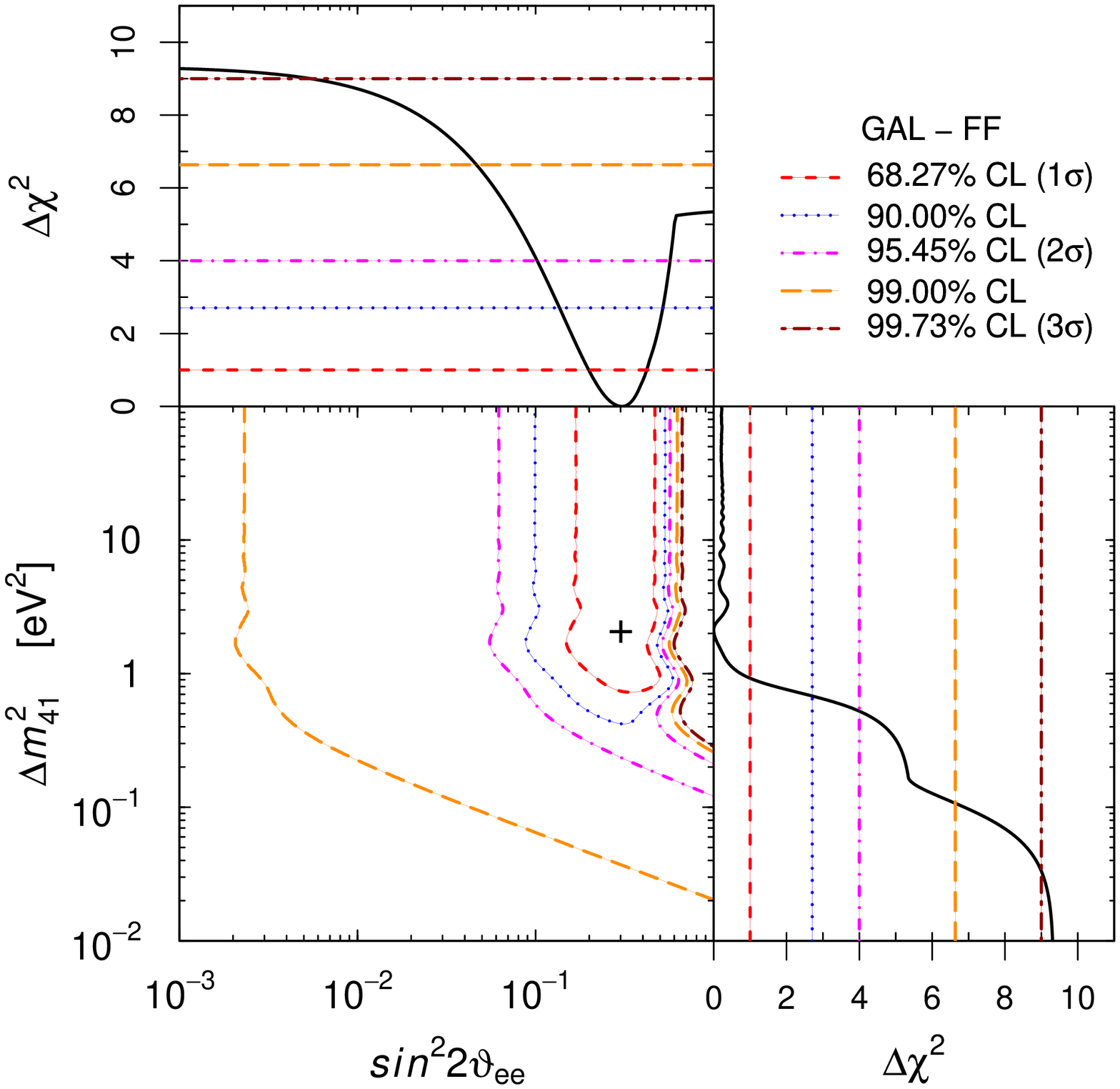}
\end{center}
\caption{ \label{fig:fit-gff}
Allowed regions and
marginal $\Delta\chi^{2}$'s
analogous to those in Fig.~\ref{fig:fit-ghk} for the FF case.
}
\end{figure}

\begin{figure}[t]
\begin{center}
\includegraphics*[width=\linewidth]{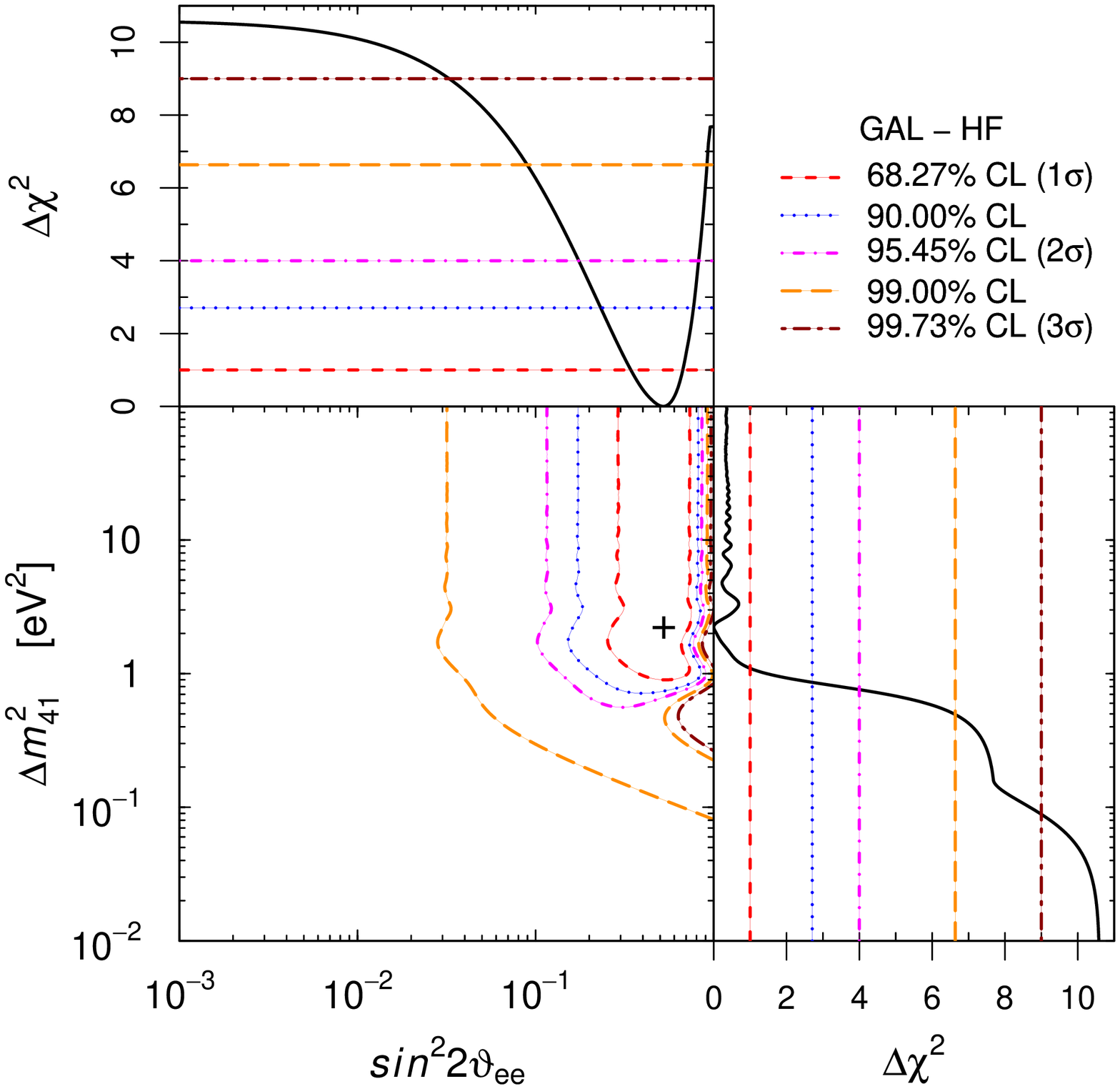}
\end{center}
\caption{ \label{fig:fit-ghf}
Allowed regions and
marginal $\Delta\chi^{2}$'s
analogous to those in Fig.~\ref{fig:fit-ghk} for the HF case.
}
\end{figure}

\begin{table}[b]
\caption{
\label{tab:fit-gal}
Values of
$\chi^{2}$,
goodness-of-fit (GoF) for 2 degrees of freedom
and
best-fit values
of the 3+1 oscillation parameters
obtained from
the three fits of
Gallium
data described in the text.
}
\begin{ruledtabular}
\begin{tabular}{cccc}
                                       &HK                                             &FF                                             &HF                                             \\
$\chi^{2}_{\text{min}}$                &4.8                &7.9                &4.6                \\
GoF                                    &9.1\%          &1.9\%          &9.9\%          \\
$\Delta{m}^2_{41}[\text{eV}^2]$        &2.24                &2.1                &2.24                \\
$\sin^22\vartheta_{ee}$                &0.50                &0.30                &0.52                \\
\end{tabular}
\end{ruledtabular}
\end{table}

We analyzed the Gallium data in the three cases above
in terms of neutrino oscillations
in the 3+1 framework,
in which the effective probability of $\nu_{e}$ survival
is given by Eq.~(\ref{survi}) with $\alpha=e$.
We used the statistical method discussed in Ref.~\cite{Giunti:2010zu},
neglecting for simplicity the small difference between the
${}^{51}\text{Cr}$ and ${}^{37}\text{Ar}$
cross section ratios in Tab.~\ref{tab:cs}.
The results of the fits are presented in Tab.~\ref{tab:fit-gal}
and Figs.~\ref{fig:fit-ghk}--\ref{fig:fit-ghf}.
One can see that in any case neutrino oscillations give an acceptable fit of the data.
In the FF case the goodness-of-fit is smaller than in the HK and HF cases, because of the much smaller uncertainty
of $\text{BGT}_{175}$.
The three cases give approximately the same best-fit value and allowed range of
$\Delta{m}^2_{41}$.
Instead, they differ in the best-fit value and allowed range of
$\sin^{2}2\vartheta_{ee}$:
the FF case is in favor of smaller values of $\sin^{2}2\vartheta_{ee}$
than the HK and HF cases.

\begin{figure}[t]
\begin{center}
\includegraphics*[width=\linewidth]{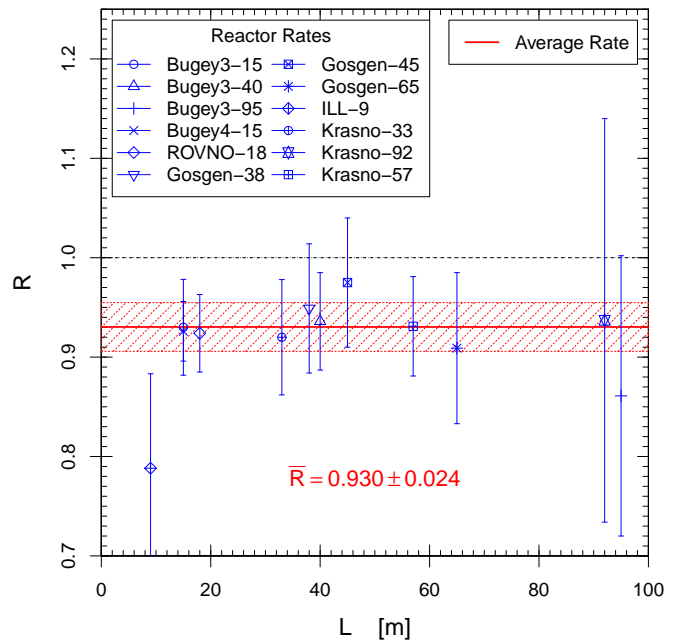}
\end{center}
\caption{ \label{fig:rea-plt}
Ratio $R$ of the observed $\bar\nu_{e}$ event rate and that expected in absence of $\bar\nu_{e}$ disappearance
in reactor neutrino experiments.
The horizontal band represents the average value of $R$ with $1\sigma$ uncertainties.
}
\end{figure}

\section{Fit of Gallium and reactor data}
\label{Fit of Gallium and reactor data}

The reactor antineutrino anomaly
\cite{Mention:2011rk}
stems from a new evaluation of the reactor $\bar\nu_{e}$ flux
\cite{Mueller:2011nm,Huber:2011wv}
which implies that the event rate measured by several reactor $\bar\nu_{e}$ experiments
at distances from the reactor core between about 10 and 100 meters
is smaller than that obtained without $\bar\nu_{e}$ disappearance.
This is illustrated in Fig.~\ref{fig:rea-plt},
where we plotted the ratio $R$ of the observed $\bar\nu_{e}$ event rate and that expected in absence of $\bar\nu_{e}$ disappearance
for the
Bugey-3 \cite{Declais:1995su},
Bugey-4 \cite{Declais:1994ma},
ROVNO91 \cite{Kuvshinnikov:1990ry},
Gosgen \cite{Zacek:1986cu},
ILL \cite{Hoummada:1995zz}
and
Krasnoyarsk \cite{Vidyakin:1990iz}
reactor antineutrino experiments.
We used the reactor neutrino fluxes presented in the recent White Paper on
light sterile neutrinos
\cite{Abazajian:2012ys},
which updates Refs.~\cite{Mueller:2011nm,Huber:2011wv,Mention:2011rk}.
From Fig.~\ref{fig:rea-plt} one can see that the reactor antineutrino anomaly has a significance of about
$2.8\sigma$.
In the fit of reactor data,
besides the above-mentioned rates, we consider also
the $40\,\text{m}/15\,\text{m}$ spectral ratio measured in the Bugey-3 experiment \cite{Declais:1995su}.

\begin{figure}[t]
\begin{center}
\includegraphics*[width=\linewidth]{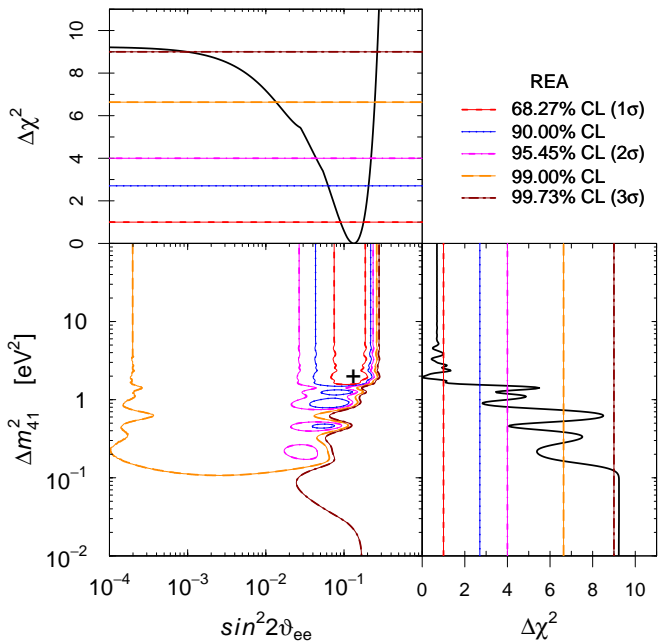}
\end{center}
\caption{ \label{fig:fit-rea}
Allowed regions in the
$\sin^{2}2\vartheta_{ee}$--$\Delta{m}^{2}_{41}$ plane
and
marginal $\Delta\chi^{2}$'s
for
$\sin^{2}2\vartheta_{ee}$ and $\Delta{m}^{2}_{41}$
obtained from the
combined fit of reactor antineutrino data.
The best-fit point corresponding to $\chi^2_{\text{min}}$ is indicated by a cross.
}
\end{figure}

\begin{table}[b]
\caption{
\label{tab:fit-rea}
Values of
$\chi^{2}$,
number of degrees of freedom (NDF),
goodness-of-fit (GoF)
and
best-fit values
of the 3+1 oscillation parameters
obtained from
the fit of reactor (REA) antineutrino data (first column)
and from the combined fit of reactor and Gallium data in the three cases discussed in Section~\ref{Gallium Anomaly}.
The last three lines give the parameter goodness-of-fit (PG)
\protect\cite{hep-ph/0304176}
of the combined fit.
}
\begin{ruledtabular}
\begin{tabular}{ccccc}
					&REA					&REA+HK						&REA+FF						&REA+HF						\\
$\chi^{2}_{\text{min}}$			&21.5	&30.6		&31.8		&31.0		\\
NDF					&36	&40		&40		&40		\\
GoF					&97\%	&86\%	&82\%	&85\%	\\
$\Delta{m}^2_{41}[\text{eV}^2]$		&1.9	&1.95		&1.95		&1.95		\\
$\sin^22\vartheta_{ee}$			&0.13	&0.16		&0.16		&0.16		\\
\hline
$\Delta\chi^{2}_{\text{PG}}$		&					&4.3	&2.4	&4.8	\\
$\text{NDF}_{\text{PG}}$		&					&2	&2	&2	\\
$\text{GoF}_{\text{PG}}$		&					&12\%	&30\%	&9\%	\\
\end{tabular}
\end{ruledtabular}
\end{table}

The results of the fit of the reactor antineutrino data
are presented in Tab.~\ref{tab:fit-rea} and Fig.~\ref{fig:fit-rea}.
One can see that the preferred range of
$\Delta{m}_{41}$
has a large overlap with that indicated by the Gallium anomaly,
but there is a strong upper bound for
$\sin^{2}2\vartheta_{ee}$
of about 0.3.
Therefore, the large-$\sin^{2}2\vartheta_{ee}$ part of the Gallium-allowed region
in each of the three cases considered in Figs.~\ref{fig:fit-ghk}--\ref{fig:fit-ghf}
is excluded by reactor data.

\begin{figure}[t]
\begin{center}
\begin{tabular}{ccc}
\includegraphics*[width=\linewidth]{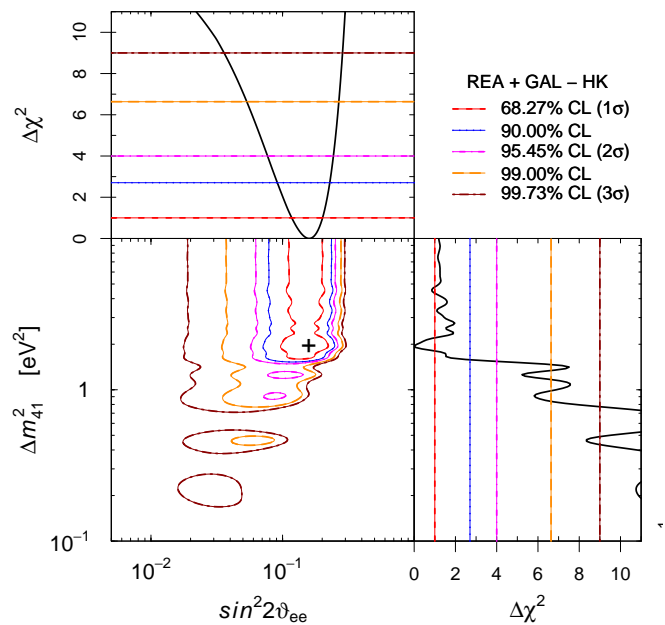}
\end{tabular}
\end{center}
\caption{ \label{fig:fit-ghk-rea}
Allowed regions in the
$\sin^{2}2\vartheta_{ee}$--$\Delta{m}^{2}_{41}$ plane
and
marginal $\Delta\chi^{2}$'s
for
$\sin^{2}2\vartheta_{ee}$ and $\Delta{m}^{2}_{41}$
obtained from the
combined fit of Gallium and reactor data
in the HF case discussed in Section~\ref{Gallium Anomaly}.
The best-fit point corresponding to $\chi^2_{\text{min}}$ is indicated by a cross.
}
\end{figure}

\begin{figure}[t]
\begin{center}
\begin{tabular}{ccc}
\includegraphics*[width=\linewidth]{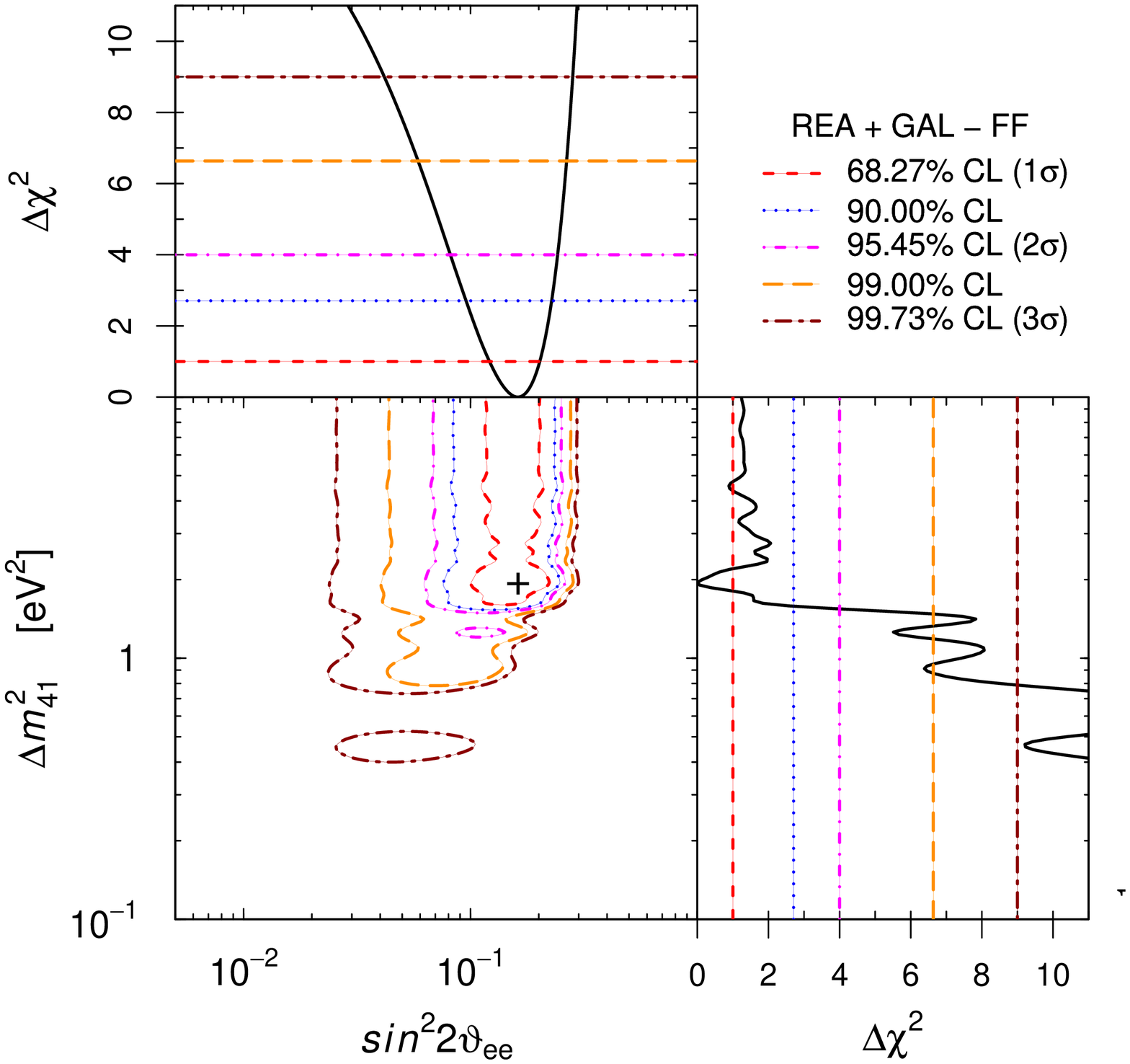}
\end{tabular}
\end{center}
\caption{ \label{fig:fit-gff-rea}
Allowed regions and
marginal $\Delta\chi^{2}$'s
analogous to those in Fig.~\ref{fig:fit-ghk-rea} for the FF case.
}
\end{figure}

\begin{figure}[t]
\begin{center}
\begin{tabular}{ccc}
\includegraphics*[width=\linewidth]{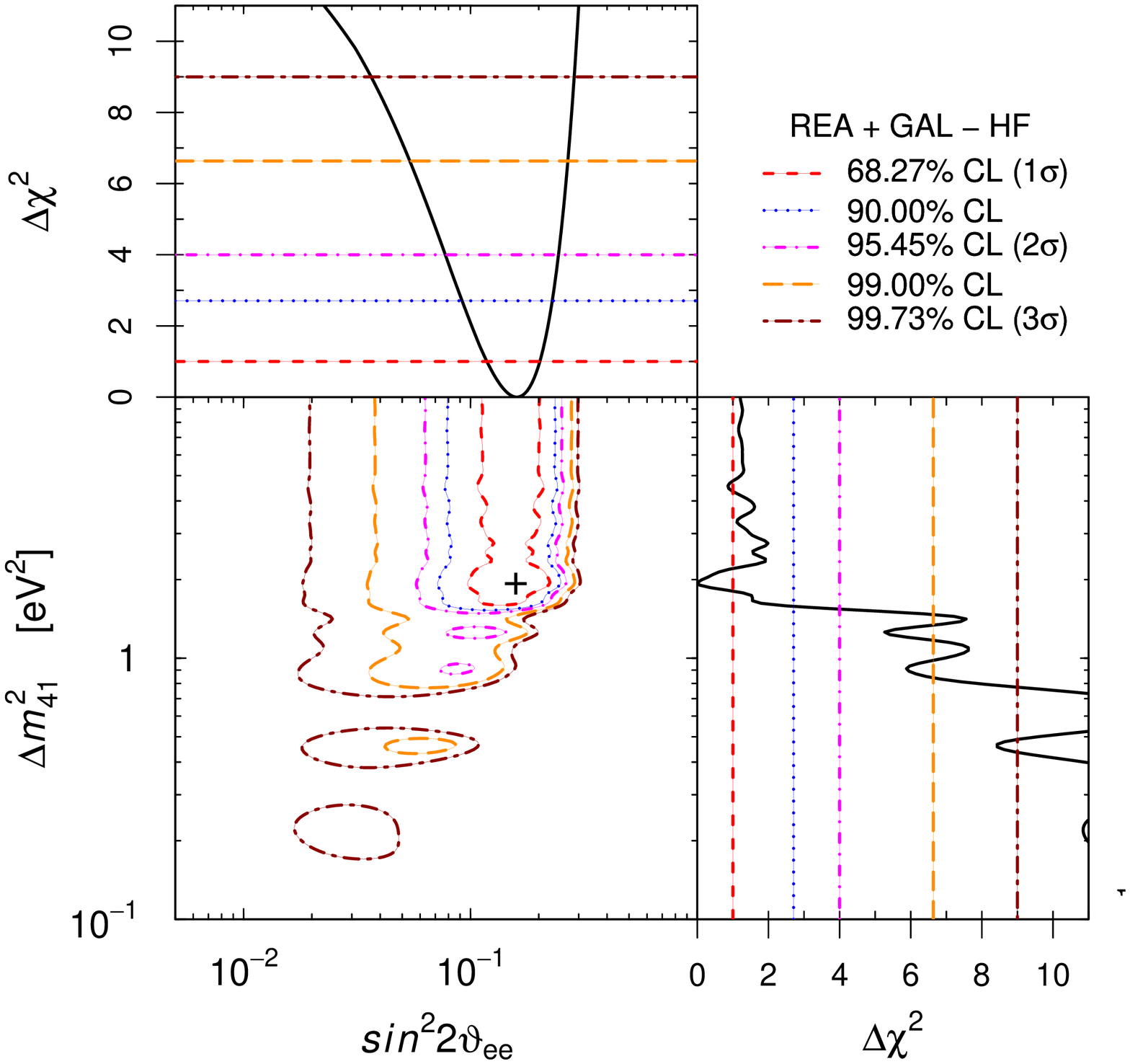}
\end{tabular}
\end{center}
\caption{ \label{fig:fit-ghf-rea}
Allowed regions and
marginal $\Delta\chi^{2}$'s
analogous to those in Fig.~\ref{fig:fit-ghk-rea} for the HF case.
}
\end{figure}

The results of the combined fit of Gallium and reactor data
are presented in Tab.~\ref{tab:fit-rea} and Figs.~\ref{fig:fit-ghk-rea}--\ref{fig:fit-ghf-rea}.
From these figures,
one can see that
the allowed regions in the
$\sin^{2}2\vartheta_{ee}$--$\Delta{m}^{2}_{41}$
in the three cases that we have considered for the fit of Gallium data
are quite similar,
and the best-fit values of
$\sin^{2}2\vartheta_{ee}$ and $\Delta{m}^{2}_{41}$
are equal
(see Tab.~\ref{tab:fit-rea}).
This is due to a dominance of reactor data, which are more numerous and have smaller uncertainties.
Hence,
in the following we consider only the FF case,
which is the one which is more compatible with reactor data,
because it agrees more than the HK and HF cases
with the reactor exclusion of large values of $\sin^{2}2\vartheta_{ee}$.
This better agreement is quantified by the larger
parameter goodness-of-fit
\cite{hep-ph/0304176}
in Tab.~\ref{tab:fit-rea}.

\section{Solar neutrino constraint}
\label{Solar neutrino constrait}

In this section we discuss the upper bound for
$\sin^{2}2\vartheta_{ee}$
which can be obtained from the
data of solar neutrino experiments
\cite{Cleveland:1998nv,1001.2731,astro-ph/0204245,hep-ex/0508053,0803.4312,1010.0118,Smy-NU2012,nucl-ex/0204009,nucl-ex/0502021,0806.0989,1104.1816,1110.3230}
and from the data of the KamLAND very-long-baseline reactor antineutrino experiment \cite{1009.4771},
which are sensitive to oscillations generated by the small squared-mass difference
$\Delta{m}^2_{21}$.
Since the event rates measured in these experiments
are well described by standard three-neutrino mixing,
the data allow us to constrain the corrections due to
active-sterile neutrino mixing,
which affects the electron neutrino and antineutrino survival probability
and generates transitions into sterile neutrinos
\cite{Dooling:1999sg,Giunti:2000wt,Giunti:2009xz,Palazzo:2011rj,Palazzo:2012yf,Palazzo-NOW2012}.
As explained in the following,
the almost degenerate effects
in the solar and KamLAND experiments
\cite{Palazzo:2011rj,Palazzo:2012yf,Palazzo-NOW2012}
of
$|U_{e4}|^2$,
which determines $\sin^{2}2\vartheta_{ee}$
through Eq.~(\ref{survisin}),
and $|U_{e3}|^2$
can be resolved by using the recent determination of
$|U_{e3}|^2$
in the
Daya Bay \cite{1203.1669}
and
RENO \cite{1204.0626}
long-baseline reactor antineutrino experiments.

The effective survival probability of electron neutrinos and antineutrinos in the
solar and KamLAND experiments
is given by\footnote{
In this discussion we neglect, for simplicity,
the matter effects in the Earth which affect the neutrino detection rates in the night,
but these effects are taken into account in our calculation.
}
\cite{Bilenky:1996rw}
\begin{equation}
P_{\nu_{e}\to\nu_{e}}^{\text{SUN}}
=
P_{\nu_{e}\to\nu_{e}}^{2\nu}
\left( 1 - \sum_{k=3}^{4} |U_{ek}|^2 \right)^2
+
\sum_{k=3}^{4} |U_{ek}|^4
\,,
\label{peesun1}
\end{equation}
where
$P_{\nu_{e}\to\nu_{e}}^{2\nu}$
is the two-neutrino survival probability
generated by $\Delta{m}^2_{21}$.
Considering small values of $|U_{e3}|^2$ and $|U_{e4}|^2$,
we have
\begin{equation}
P_{\nu_{e}\to\nu_{e}}^{\text{SUN}}
\simeq
P_{\nu_{e}\to\nu_{e}}^{2\nu}
\left( 1 - 2 \sum_{k=3}^{4} |U_{ek}|^2 \right)
\,.
\label{peesun2}
\end{equation}
In vacuum
the two-neutrino survival probability
$P_{\nu_{e}\to\nu_{e}}^{2\nu}$
has the standard two-neutrino form
which does not depend on
$|U_{e3}|^2$ and $|U_{e4}|^2$.
Therefore,
in the KamLAND experiment
$|U_{e3}|^2$ and $|U_{e4}|^2$
have the same effect of suppressing the electron neutrino and antineutrino survival probability.
For solar neutrinos,
the main effect of
$|U_{e3}|^2$ and $|U_{e4}|^2$
is the same as in the vacuum case,
but there are corrections caused by the
modifications of
$P_{\nu_{e}\to\nu_{e}}^{2\nu}$
due to
the decrease of
$|U_{e1}|^2 + |U_{e2}|^2 = 1 - ( |U_{e3}|^2 + |U_{e4}|^2 )$
if
$|U_{e3}|^2\neq0$ and/or $|U_{e4}|^2\neq0$
and to the contribution of the neutral current potential $V_{\text{NC}}$
which is not felt by the sterile neutrino
during propagation in matter.

In order to describe this effect,
we neglect possible CP-violating phases in the mixing matrix
and
we parameterize it as
(see also \cite{Palazzo:2011rj})
\begin{equation}
U = R_{23} R_{24} R_{34} R_{14} R_{13} R_{12}
\,,
\label{urot}
\end{equation}
where
$R_{ab}$ is the real orthogonal matrix
($R_{ab}^{T}=R_{ab}^{-1}$)
which operates a rotation in the $a$-$b$ plane by an angle $\vartheta_{ab}$:
\begin{align}
[R^{ab}]_{rs}
=
\delta_{rs}
\null & \null
+
\left( c_{ab} - 1 \right)
\left( \delta_{ra} \, \delta_{sa} + \delta_{rb} \, \delta_{sb} \right)
\nonumber
\\
\null & \null
+
s_{ab}
\left(
\delta_{ra} \, \delta_{sb}
-
\delta_{rb} \, \delta_{sa}
\right)
\,,
\label{Rab}
\end{align}
with
$c_{ab} \equiv \cos\vartheta_{ab}$
and
$s_{ab} \equiv \sin\vartheta_{ab}$.
In this parameterization
the electron line of the mixing matrix is given by
\begin{align}
\null & \null
U_{e1} = c_{12} c_{13} c_{14}
\,,
\quad
\null && \null
U_{e2} = s_{12} c_{13} c_{14}
\,,
\label{Ue12}
\\
\null & \null
U_{e3} = s_{13} c_{14}
\,,
\quad
\null && \null
U_{e4} = s_{14}
\,.
\label{Ue34}
\end{align}
Hence,
this is an extension of the standard three-neutrino mixing parameterization of the electron line
(see \cite{Giunti:2007ry,Bilenky:2010zza,Xing:2011zza})
with the addition of $\nu_{e}$-$\nu_{4}$ mixing parameterized by $\vartheta_{14}$.
This parameterization is also convenient because
\begin{equation}
\vartheta_{ee} = \vartheta_{14}
\,.
\label{tee14}
\end{equation}

Since the sterile line of the mixing matrix is more complicated,
it is convenient to write its first two elements as
\begin{equation}
U_{s1} = \cos\varphi_{s} \cos\chi_{s}
\,,
\quad
U_{s2} = \sin\varphi_{s} \cos\chi_{s}
\,,
\label{Us12}
\end{equation}
with
\begin{align}
\null & \null
\tan\varphi_{s}
=
\frac{Z s_{12} c_{24} + c_{12} s_{24}}{Z c_{12} c_{24} - s_{12} s_{24}}
\,,
\label{tanphis}
\\
\null & \null
\cos^{2}\chi_{s}
=
1 - \sum_{k=3}^{4} |U_{sk}|^2
=
Z^2 c_{24}^2 + s_{24}^2
\,,
\label{coschis}
\\
\null & \null
Z = c_{13} c_{34} s_{14} - s_{13} s_{34}
\,.
\label{Zste}
\end{align}

The adiabatic two-neutrino survival probability $P_{\nu_{e}\to\nu_{e}}^{2\nu}$ is given by
\cite{Giunti:2009xz}
\begin{equation}
P_{\nu_{e}\to\nu_{e}}^{2\nu}
=
\frac{1}{2}
\left(
1
+
\cos2\vartheta_{12}
\cos2\vartheta_{12}^{0}
\right)
\,,
\label{pee2nu}
\end{equation}
where
$\vartheta_{12}^{0}$ is the effective mixing angle at neutrino production,
which is given by
\begin{equation}
\vartheta_{12}^{0} = \vartheta_{12} + \omega^{0}
\,,
\label{t120}
\end{equation}
The mixing angle $\omega$
between the vacuum mass basis and the effective mass basis in matter is given by
\begin{equation}
\tan2\omega
=
\frac{ 2 E V \sin 2 \xi }{ \Delta{m}^2_{21} - 2 E V \cos 2 \xi }
\,.
\label{043}
\end{equation}
Here $V$ is the matter potential given by
\begin{align}
\null & \null
V^2
=
V_{\text{CC}}^2 c_{13}^4 c_{14}^4
+
V_{\text{NC}}^2 \cos^{4}\chi_{s}
\nonumber
\\
\null & \null
-
2 V_{\text{CC}} V_{\text{NC}} \cos2(\vartheta_{12}-\varphi_{s}) c_{13}^2 c_{14}^2 \cos^{2}\chi_{s}
\,,
\label{matpot}
\end{align}
where $V_{\text{CC}}$ and $V_{\text{NC}}$
are the standard charged-current and neutral-current matter potentials.
The angle $\xi$
is given by
\begin{equation}
\tan2\xi
=
\frac
{V_{\text{CC}} \sin2\vartheta_{12} c_{13}^2 c_{14}^2 - V_{\text{NC}} \sin2\varphi_{s} \cos^{2}\chi_{s}}
{V_{\text{CC}} \cos2\vartheta_{12} c_{13}^2 c_{14}^2 - V_{\text{NC}} \cos2\varphi_{s} \cos^{2}\chi_{s}}
\,.
\label{xi}
\end{equation}
Therefore,
for $s_{14}\ll1$
the contributions of
$|U_{e3}|^2 \simeq s_{13}^2$
and $|U_{e4}|^2=s_{14}^2$
to the matter effects
are almost degenerate.
There is only a small difference of their contributions
due to $Z$ in Eq.~(\ref{Zste}).

In solar neutrino measurements the degeneracy
of the effects of
$|U_{e3}|^2$ and $|U_{e4}|^2$
is also slightly broken by the SNO neutral current measurement,
which is sensitive to the total probability of $\nu_{e}$ transitions into active neutrinos,
which by unitarity is given by
$1 - P_{\nu_{e}\to\nu_{s}}^{\text{SUN}}$,
with
\begin{equation}
P_{\nu_{e}\to\nu_{s}}^{\text{SUN}}
=
P_{\nu_{e}\to\nu_{s}}^{2\nu}
c_{13}^2 c_{14}^2
\cos^2\chi_{s}
+
\sum_{k=3}^{4} |U_{ek}|^2 |U_{sk}|^2
\,.
\label{pes-sun}
\end{equation}
Here,
the adiabatic two-neutrino transition probability $P_{\nu_{e}\to\nu_{s}}^{2\nu}$
is given by\footnote{
\label{fot1}
One can check that
in the limit of two-neutrino $\nu_{e}$-$\nu_{s}$ mixing
the unitarity relation
$P_{\nu_{e}\to\nu_{e}}^{2\nu} + P_{\nu_{e}\to\nu_{s}}^{2\nu} = 1$
is satisfied.
In this case,
$c_{13} = c_{14} = s_{24} = \cos\chi_{s} = 1$
and
$\varphi_{s} = \vartheta_{12} + \pi/2$.
}
\cite{Giunti:2009xz}
\begin{equation}
P_{\nu_{e}\to\nu_{s}}^{2\nu}
=
\frac{1}{2}
\left(
1
+
\cos2\varphi_{s}
\cos2\vartheta_{12}^{0}
\right)
\,.
\label{pes2nu}
\end{equation}
The degeneracy of $|U_{e3}|^2$ and $|U_{e4}|^2$ is broken by their different effects in
$\varphi_{s}$,
$\chi_{s}$
and in the last term of Eq.~(\ref{pes-sun}).

Luckily,
the recent determination of the value of $\vartheta_{13}$
in the Daya Bay \cite{1203.1669} and
RENO \cite{1204.0626} experiment allow us to
obtain information on the value of
$\vartheta_{ee} = \vartheta_{14}$
without much uncertainty due to $\vartheta_{13}$.
In the 3+1 mixing scheme under consideration
the effective long-baseline $\bar\nu_{e}$ survival probability
in the Daya Bay and RENO far detectors is given by
\cite{0809.5076}
\begin{equation}
P_{\bar\nu_{e}\to\bar\nu_{e}}^{\text{LBL-F}}
=
1
-
c^4_{14} \sin^2 2 \vartheta_{13}
\sin^2\left( \frac{\Delta{m}^2_{31}L}{4E} \right)
-
\frac{1}{2}
\sin^2 2 \vartheta_{14}
\,,
\label{PeeLBLfar}
\end{equation}
since the oscillations due to
$\Delta{m}^2_{41}\gg\Delta{m}^2_{31}$
are averaged
and the oscillations due to
$\Delta{m}^2_{21}\ll\Delta{m}^2_{31}$
are negligibly small.
This survival probability depends on $\vartheta_{14}$,
but the Daya Bay and RENO collaborations
measured the ratio of the probability (\ref{PeeLBLfar}) in the far detectors
and
the probability (\ref{PeeLBLfar}) with
$\Delta{m}^2_{31}L/4E \ll 1$
in the near detectors,
\begin{equation}
P_{\bar\nu_{e}\to\bar\nu_{e}}^{\text{LBL-N}}
=
1
-
\frac{1}{2}
\sin^2 2 \vartheta_{14}
\,.
\label{PeeLBLnear}
\end{equation}
Since the contribution of small values of
$\vartheta_{14}$
to the measured ratio is of order
$\vartheta_{14}^4$
\cite{Giunti:2011vc},
in practice
the value of
$\vartheta_{13}$
determined by the Daya Bay and RENO collaborations
with a three-neutrino mixing survival probability
is accurate also in the 3+1 scheme under consideration.
Nevertheless,
since we have the possibility,
in our analysis we took into account the exact ratio of the far and near survival probabilities
(\ref{PeeLBLfar}) and (\ref{PeeLBLnear})
by including the least-squares function
$\chi^2_{\text{LBL}}$
of the far/near relative measurements of Daya Bay and RENO in the total solar and reactor
least-squares function
\begin{equation}
\chi^2
=
\chi^2_{\text{SOL}}
+
\chi^2_{\text{KL}}
+
\chi^2_{\text{LBL}}
\,.
\label{chi2sun}
\end{equation}

For the calculation of the solar least-squares function
$\chi^2_{\text{SOL}}$, we considered the radiochemical
$^{\text{37}}\text{Cl}$ \cite{Cleveland:1998nv} and $^{\text{71}}\text{Ga}$
\cite{1001.2731,astro-ph/0204245} experiments, the day and night
energy spectra of all four phases of Super-Kamiokande
\cite{hep-ex/0508053,0803.4312,1010.0118,Smy-NU2012}, the day and
night energy spectra of the SNO $\text{D}_2\text{O}$ phase
\cite{nucl-ex/0204009}, the charged-current and neutral-current
rates of the SNO salt \cite{nucl-ex/0502021} and NCD
\cite{0806.0989} phases, and the rates of low energy
$^{\text{7}}\text{Be}$ \cite{1104.1816} and pep \cite{1110.3230} solar neutrinos from
the Borexino experiment.
We did not use the SNO data obtained with the low energy threshold analysis \cite{0910.2984}
and those obtained with the combined analysis \cite{1109.0763} because both analyses
assumed a three-parameter polynomial survival probability which
is not appropriate for the sterile neutrino analysis.
The solar neutrino fluxes are taken from the BP2004 \cite{Bahcall:2004fg}
standard solar model,
except for the normalization of the
solar $^8\text{B}$ neutrino flux,
which is considered as a free parameter determined by the minimization of $\chi^2$
(as usual, because of its large theoretical uncertainties).

The KamLAND least-squares function
$\chi^2_{\text{KL}}$
has been calculated using the energy spectrum reported in \cite{1009.4771} with a total
exposure of $3.49\times 10^{32}$ target-proton-year.

In our analysis, we used the $\nu_{e}$ survival probability
(\ref{peesun1})
and the $\nu_e\to\nu_s$ transition probability
(\ref{pes-sun})
taking into account as parameters
the squared-mass difference
$\Delta{m}^{2}_{21}$
and the five relevant mixing angles
$\vartheta_{12}$,
$\vartheta_{13}$,
$\vartheta_{14}$,
$\vartheta_{24}$,
$\vartheta_{34}$
(solar and KamLAND oscillations are independent from
$\vartheta_{23}$,
because $\nu_{\mu}$ and $\nu_{\tau}$
are indistinguishable;
see \cite{Giunti:2007ry}).
The six-dimensional parameter space is explored
with a Markov Chain Monte Carlo sampling
in order to minimize the total $\chi^2$ in Eq.~(\ref{chi2sun}).

\begin{figure}[t]
\begin{center}
\includegraphics*[width=\linewidth]{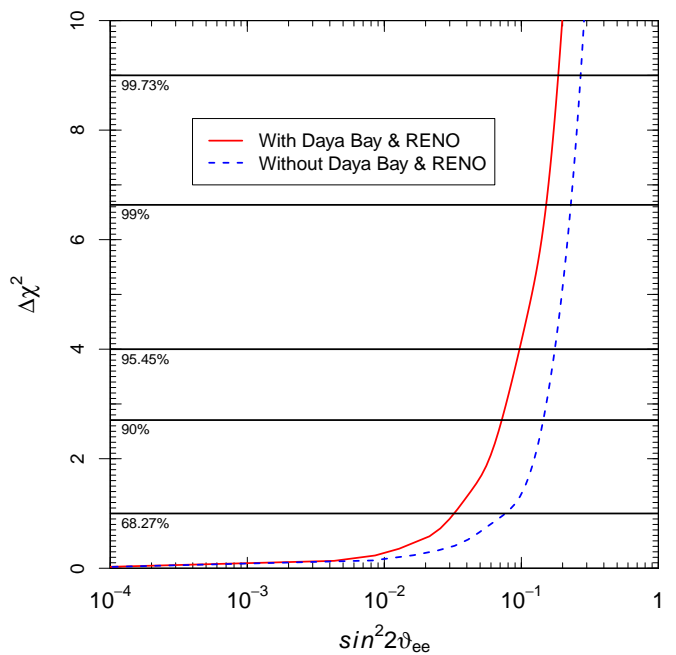}
\end{center}
\caption{ \label{fig:sun}
Marginalized $\Delta\chi^2 = \chi^2 - \chi^2_{\text{min}}$
as a function of
$\sin^{2}2\vartheta_{ee}$
obtained from the fit of
solar and KamLAND data
with and without
Daya Bay and RENO data.
}
\end{figure}

Since the best fit is obtained for
$\vartheta_{14}=0$,
the marginalized $\Delta\chi^2 = \chi^2 - \chi^2_{\text{min}}$
shown in Fig.~\ref{fig:sun}
give stringent constraints on the value of
$\sin^{2}2\vartheta_{ee} = \sin^{2}2\vartheta_{14}$.
In Fig.~\ref{fig:sun} we have plotted the $\Delta\chi^2$
obtained with and without including
the Daya Bay and RENO data.
One can see that these data are useful in order to tighten the upper bound on $\sin^{2}2\vartheta_{ee}$.

\section{Global fit}
\label{Global fit}

In this section we present the results of the global fit of
electron neutrino and antineutrino disappearance data,
which includes the Gallium and reactor data discussed respectively in
Sections~\ref{Gallium Anomaly} and \ref{Fit of Gallium and reactor data},
the solar neutrino constraint discussed in Section~\ref{Solar neutrino constrait},
and the
KARMEN \cite{Bodmann:1994py,hep-ex/9801007}
and
LSND \cite{hep-ex/0105068}
$\nu_{e} + {}^{12}\text{C} \to {}^{12}\text{N}_{\text{g.s.}} + e^{-}$
scattering data \cite{1106.5552},
with the method discussed in Ref.~\cite{Giunti:2011cp}.

\begin{figure}[t]
\begin{center}
\includegraphics*[width=\linewidth]{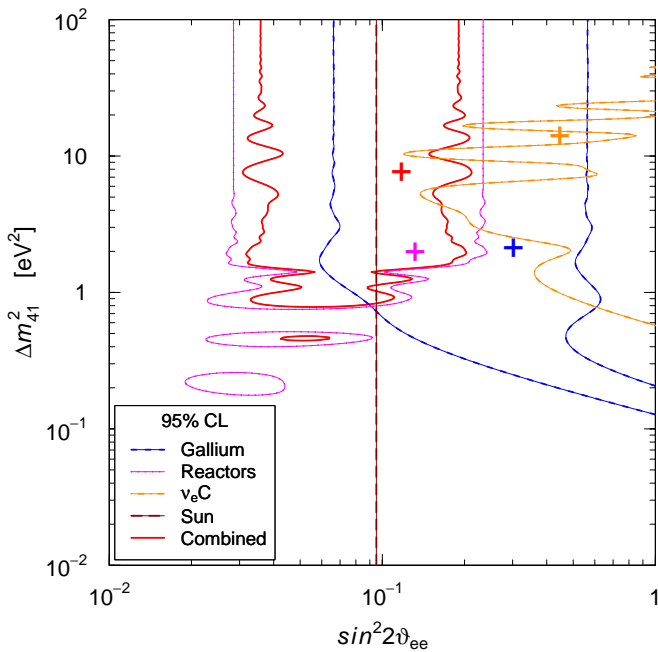}
\end{center}
\caption{ \label{fig:fit-glo-sup}
Allowed 95\% CL regions in the
$\sin^{2}2\vartheta_{ee}$--$\Delta{m}^{2}_{41}$ plane
obtained from the separate fits of Gallium, reactor, solar and $\nu_{e}\text{C}$ scattering data
and from the combined fit of all data.
The best-fit points corresponding to $\chi^2_{\text{min}}$ are indicated by crosses.
}
\end{figure}

\begin{figure}[t]
\begin{center}
\includegraphics*[width=\linewidth]{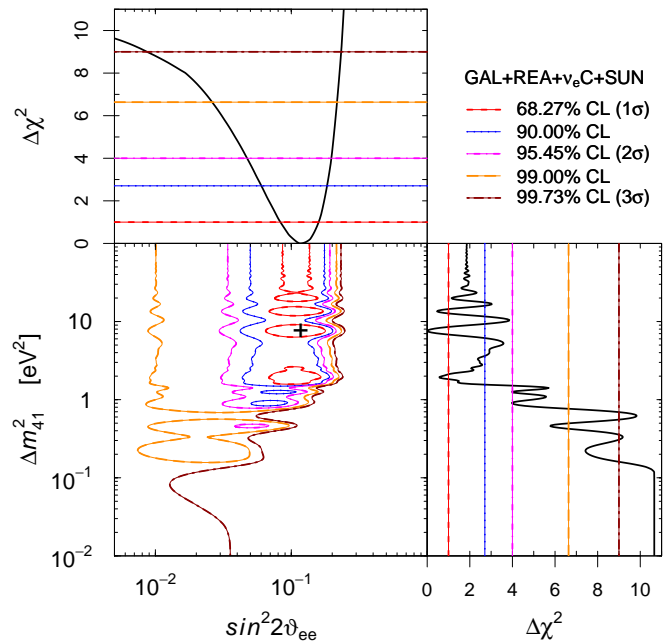}
\end{center}
\caption{ \label{fig:fit-glo}
Allowed regions in the
$\sin^{2}2\vartheta_{ee}$--$\Delta{m}^{2}_{41}$ plane
and
marginal $\Delta\chi^{2}$'s
for
$\sin^{2}2\vartheta_{ee}$ and $\Delta{m}^{2}_{41}$
obtained from the global fit of $\nu_{e}$ and $\bar\nu_{e}$ data.
The best-fit point corresponding to $\chi^2_{\text{min}}$ is indicated by a cross.
}
\end{figure}

Figure~\ref{fig:fit-glo-sup} shows a comparison of the
allowed 95\% CL regions in the
$\sin^{2}2\vartheta_{ee}$--$\Delta{m}^{2}_{41}$ plane
obtained from the separate fits of Gallium, reactor, solar and $\nu_{e}\text{C}$ scattering data
and from the
combined fit of all data.
One can see that the separate allowed regions overlap in a band
delimited by $\Delta{m}^{2}_{41} \gtrsim 1\,\text{eV}^2$
and
$0.07 \lesssim \sin^{2}2\vartheta_{ee} \lesssim 0.09$,
which is included in the globally allowed 95\% CL region.
Figure~\ref{fig:fit-glo} shows the globally allowed
regions in the
$\sin^{2}2\vartheta_{ee}$--$\Delta{m}^{2}_{41}$ plane and the
marginal $\Delta\chi^{2}$'s
for
the two oscillation parameters.
The best-fit point is at a relatively large value of
$\Delta{m}^{2}_{41}$,
\begin{equation}
(\Delta{m}^{2}_{41})_{\text{bf}}
=
7.6
\,
\text{eV}^2
\,,
\qquad
(\sin^{2}2\vartheta_{ee})_{\text{bf}}
=
0.12
\,,
\label{bf}
\end{equation}
with
$
\chi^2_{\text{min}}/\text{NDF}
=
45.5
/
51
$,
corresponding to a
$69$\%
goodness-of-fit.
However, there is a region allowed at $1\sigma$
around
$\Delta{m}^{2}_{41} \simeq 2 \, \text{eV}^2$
and
$\sin^{2}2\vartheta_{ee} \simeq 0.1$.
The slight preference of the global fit for
$\Delta{m}^{2}_{41} \simeq 7.6 \, \text{eV}^2$
with respect to
$\Delta{m}^{2}_{41} \simeq 2 \, \text{eV}^2$
(see the marginal $\Delta\chi^{2}$ for $\Delta{m}^{2}_{41}$ in Fig.~\ref{fig:fit-glo}),
which is preferred by Gallium and reactor data
(see Tabs.~\ref{tab:fit-gal} and \ref{tab:fit-rea}
and Figs.~\ref{fig:fit-ghk}--\ref{fig:fit-ghf}, \ref{fig:fit-rea} and \ref{fig:fit-ghk-rea}--\ref{fig:fit-ghf-rea}),
is due to the
$\nu_{e}\text{C}$ scattering data,
which prefer larger values of $\Delta{m}^{2}_{41}$
(see the discussion in Ref.~\cite{Giunti:2011cp}).

Comparing the minimum of the $\chi^2$ of the global fit
with the sum of the minima of the $\chi^2$
of the separate fits of
Gallium, reactor, solar and
$\nu_{e}\text{C}$
scattering data,
we obtained
$\Delta\chi^{2}_{\text{PG}} = 11.5$,
with
$5$
degrees of freedom,
which gives a parameter goodness-of-fit of
4\%.
Therefore, the compatibility of the four data sets is acceptable.

\begin{figure}[t]
\begin{center}
\includegraphics*[width=\linewidth]{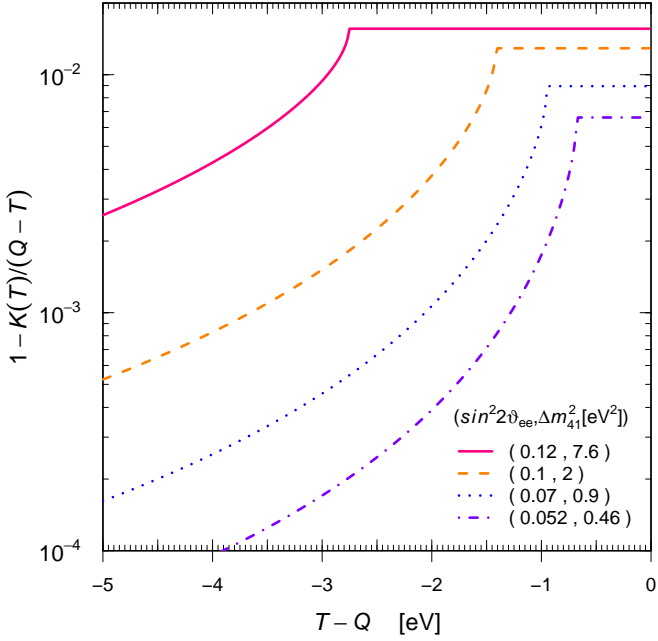}
\end{center}
\caption{ \label{fig:kur}
Relative deviation of the Kurie plot in $\beta$-decay
for some points in the allowed regions
of Fig.~\ref{fig:fit-glo}.
}
\end{figure}

The results of the global fit, as well as the results of the fits of Gallium and reactor data,
lead to lower limits for
$\Delta{m}^2_{41}$,
but there is no upper limit for
$\Delta{m}^2_{41}$
in Figs.~\ref{fig:fit-ghk}--\ref{fig:fit-ghf}, \ref{fig:fit-rea}, \ref{fig:fit-ghk-rea}--\ref{fig:fit-ghf-rea} and \ref{fig:fit-glo}.
Hence, one can ask if there are other measurements which constrain large values of
$\Delta{m}^2_{41}$.
The answer is positive and comes from the
measurements of the effects of heavy neutrino masses
on electron spectrum in $\beta$-decay far from the end-point,
from the results of neutrinoless double-$\beta$ decay experiments
for Majorana neutrinos,
and from cosmological measurements.
In this discussion we consider the mass hierarchy
\begin{equation}
m_{4} \gg m_{1},m_{2},m_{3}
\,,
\label{mass-hierarchy}
\end{equation}
which implies
\begin{equation}
m_{4} \simeq \sqrt{\Delta{m}^2_{41}}
\,.
\label{m4}
\end{equation}

Let us consider first $\beta$-decay experiments.
The ratio of the Kurie function $K(T)$ in $\beta$-decay for the case of a heavy neutrino $\nu_{4}$
and that corresponding to massless neutrinos is given by
\cite{Giunti:2011cp}
\begin{align}
\null & \null
\left( \frac{K(T)}{Q-T} \right)^2
=
1 - |U_{e4}|^2
\nonumber
\\
\null & \null
+ |U_{e4}|^2 \sqrt{1-\frac{m_{4}^2}{(Q-T)^2}} \, \theta(Q-T-m_{4})
\,,
\label{kur}
\end{align}
where
$T$ is the kinetic energy of the electron,
$Q=18.574\,\text{keV}$
is the $Q$-value of the decay,
$\theta$ is the Heaviside step function,
and we have neglected the contribution of the three light neutrinos
$\nu_{1}$,
$\nu_{2}$,
$\nu_{3}$.
Figure~\ref{fig:kur} shows the relative deviation of the Kurie plot
with respect to the massless case
for some points in the allowed regions
of Fig.~\ref{fig:fit-glo}.
One can see that in order to see the effect of $m_{4}$,
beta-decay experiments must have a sensitivity to the relative deviation of the Kurie plot
of the order of a percent or better for
$T \gtrsim Q-m_{4}$.

In 2001 the Genoa $^{187}\text{Re}$ beta-decay experiment \cite{Galeazzi:2001py}
searched for
deviation of the electron spectrum due to a heavy neutrino with a mass from 50 to 1000 eV.
From Fig.~3 of Ref.~\cite{Galeazzi:2001py}
one can see that the 95\% CL upper bound for $m_{4}$ is about 300 eV
if $\sin^{2}2\vartheta_{ee}\simeq0.1$,
which implies a very large upper limit
on $\Delta{m}^2_{41}$ of about $10^{5}\,\text{eV}^2$.

\begin{figure}[t]
\begin{center}
\includegraphics*[width=\linewidth]{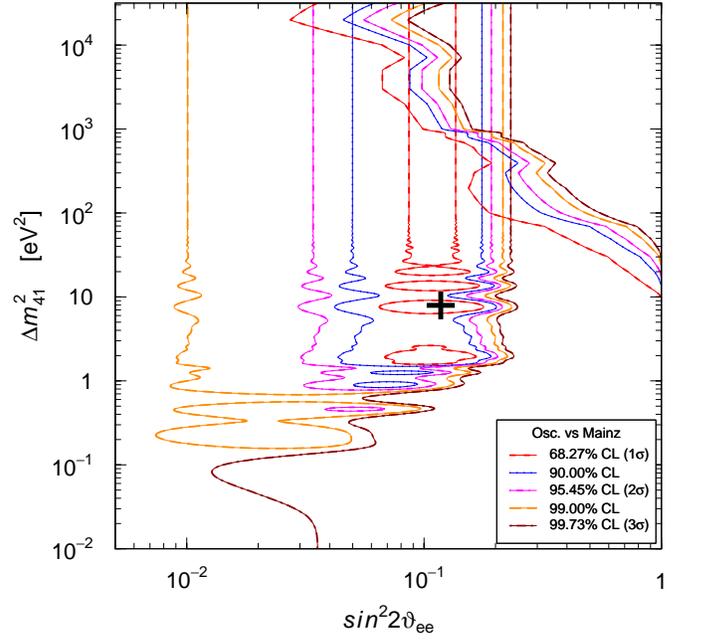}
\end{center}
\caption{ \label{fig:con-mnz}
Comparison of the
Mainz $\beta$-decay bound
(curves in the top-right part of the figure)
with the
allowed regions in the
$\sin^{2}2\vartheta_{ee}$--$\Delta{m}^{2}_{41}$ plane
obtained from the global fit of $\nu_{e}$ disappearance data
(same as in Fig.~\ref{fig:fit-glo}).
}
\end{figure}

\begin{figure}[t]
\begin{center}
\includegraphics*[width=\linewidth]{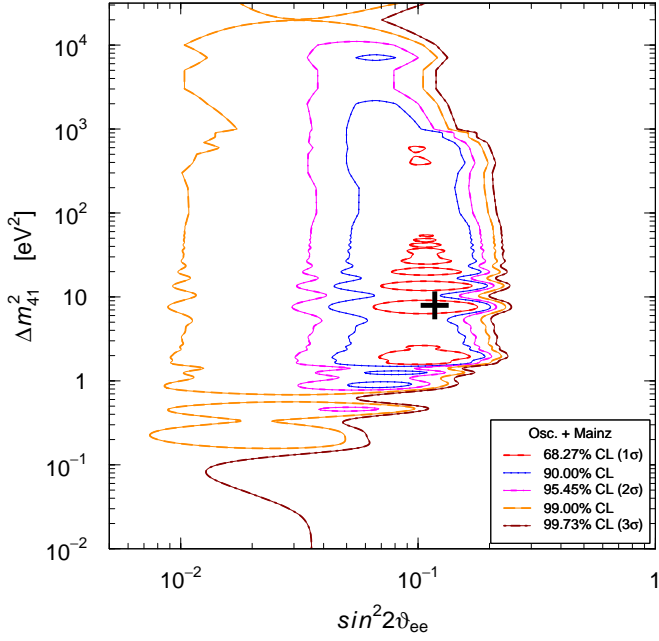}
\end{center}
\caption{ \label{fig:con-osc-mnz}
Allowed regions in the
$\sin^{2}2\vartheta_{ee}$--$\Delta{m}^{2}_{41}$ plane
obtained from the combined fit of
$\nu_{e}$ disappearance and Mainz $\beta$-decay data.
}
\end{figure}

Very recently,
the Mainz collaboration released new data
obtained with the phase II of the Mainz Neutrino Mass Experiment
\cite{1210.4194}
which constrain the value of
$\sin^{2}\vartheta_{ee}$
for $m_{4}^2$
between about 10 and $3\times10^{4} \, \text{eV}^2$.
Figure~\ref{fig:con-mnz} shows
the constraints in the
$\sin^{2}2\vartheta_{ee}$--$\Delta{m}^{2}_{41}$ plane
that we obtained with a $\chi^2$ analysis of the Mainz data in \cite{1210.4194}.
From the comparison with the allowed regions
obtained from the global fit of $\nu_{e}$ disappearance data
shown in Fig.~\ref{fig:con-mnz}
one can see that the Mainz data constrain
$\Delta{m}^{2}_{41}$
to be smaller than about $10^{4}\,\text{eV}^2$
at about 90\% CL.
This is confirmed by the results of the
combined fit shown in Fig.~\ref{fig:con-osc-mnz}.

The KATRIN experiment
(see \cite{Otten:2008zz}),
which will start in 2015 \cite{Wolf-NPB2012},
may be able to improve dramatically the upper limits
on $m_{4}$
and maybe see its effects on the electron spectrum
\cite{Esmaili:2012vg}.

\begin{figure}[t]
\begin{center}
\includegraphics*[width=\linewidth]{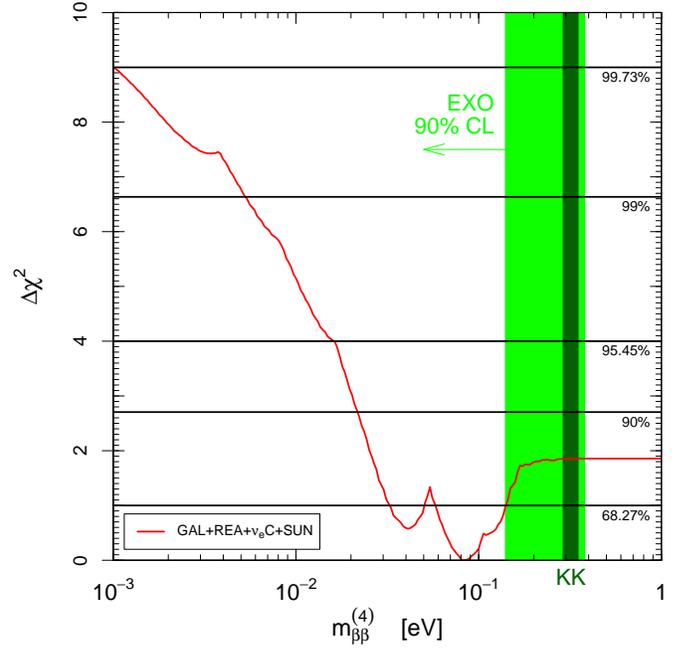}
\end{center}
\caption{ \label{fig:mbb-chi}
Marginal
$\Delta\chi^2 = \chi^2 - \chi^2_{\text{min}}$
as a function of
$m_{\beta\beta}^{(4)}$
obtained from the global fit.
The vertical green band represents the currently most stringent 90\% CL upper bound for $m_{\beta\beta}^{(4)}$
in the no-cancellation case (\ref{no-cancellation})
obtained from the 90\% CL EXO bound on $m_{\beta\beta}$ taking into account nuclear matrix element uncertainties
\cite{Auger:2012ar}.
The vertical dark-green band corresponds to the $1\sigma$
Klapdor-Kleingrothaus et al. range of $m_{\beta\beta}$
\cite{KlapdorKleingrothaus:2006ff}.
}
\end{figure}

\begin{figure*}[t]
\null
\hfill
\includegraphics*[width=0.49\linewidth]{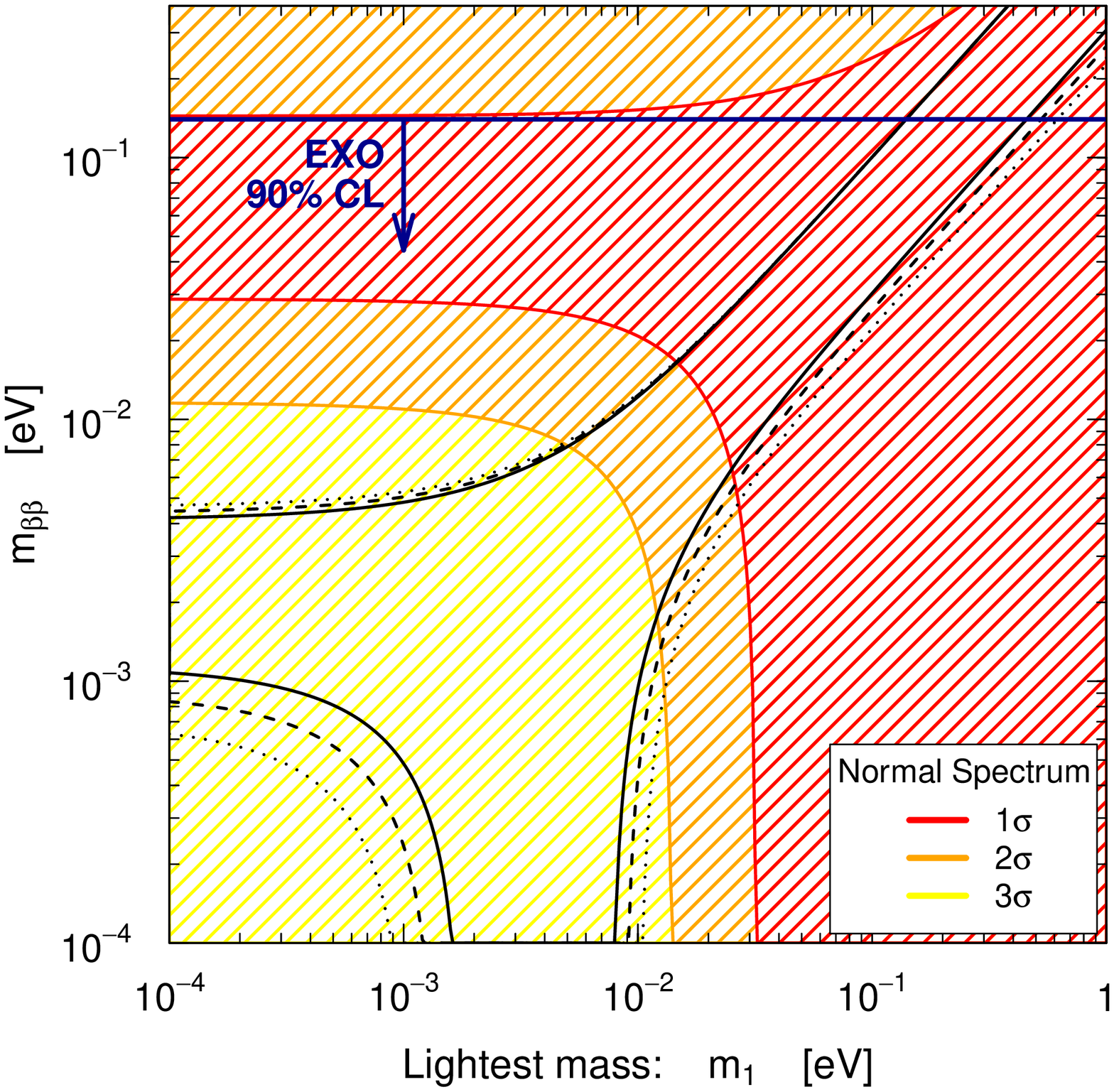}
\hfill
\includegraphics*[width=0.49\linewidth]{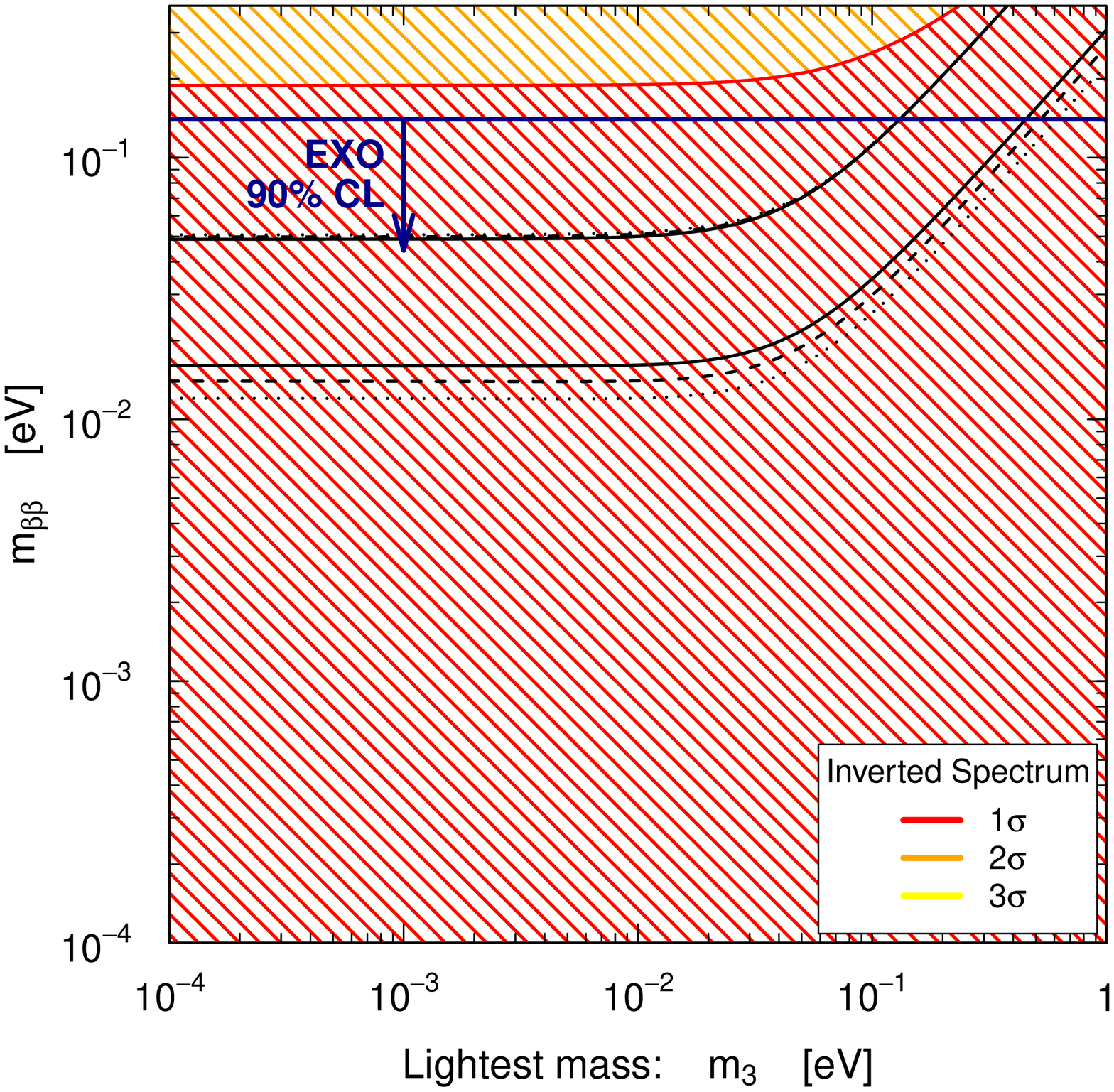}
\hfill
\null
\caption{ \label{fig:mbb}
Allowed ranges of the effective Majorana mass $|m_{\beta\beta}|$ as a
function of the lightest neutrino mass
in the two 3+1 schemes
with a normal (left)
and
inverted (right)
mass spectra of the three light neutrinos
$\nu_{1}$,
$\nu_{2}$,
$\nu_{3}$.
The black lines delimit the three-neutrino mixing
allowed regions at
$1\sigma$ (solid),
$2\sigma$ (dashed),
$3\sigma$ (dotted)
\cite{1203.5250}.
}
\end{figure*}

The heavy neutrino mass $m_{4}$ has also an effect in neutrinoless double-$\beta$ decay
(see \cite{1109.5515,1203.5250,1205.0649,1206.2560}),
if massive neutrinos are Majorana particles
(see \cite{Giunti:2007ry,Bilenky:2010zza,Xing:2011zza}).
Considering Eq.~(\ref{m4}),
the contribution of the heavy neutrino mass $m_{4}$
to the effective Majorana mass
\begin{equation}
m_{\beta\beta}
=
\left| \sum_{k} U_{ek}^2 m_{k} \right|
\label{majorana}
\end{equation}
is given by
\begin{equation}
m_{\beta\beta}^{(4)}
\simeq
|U_{e4}|^2 \sqrt{\Delta{m}^2_{41}}
\,.
\label{m2b4}
\end{equation}
Figure~\ref{fig:mbb-chi} shows the marginal
$\Delta\chi^2 = \chi^2 - \chi^2_{\text{min}}$
as a function of $m_{\beta\beta}^{(4)}$
obtained from the global fit.
One can see that $m_{\beta\beta}^{(4)}$ is bounded from below
and it is likely to be larger than about $10^{-2} \, \text{eV}$,
a value which may be reached in the next generation of
neutrinoless double-$\beta$ decay experiment
(see \cite{1206.2560,Cremonesi-NOW2012}).
Of course,
if the three light neutrinos
$\nu_{1}$,
$\nu_{2}$,
$\nu_{3}$
are quasi-degenerate at a mass scale larger than about $10^{-2} \, \text{eV}$
the contribution of $m_{\beta\beta}^{(4)}$
can cancel with that of the three light neutrinos.
Such an unfortunate cancellation can also happen if
the masses of the three light neutrinos follow an inverted hierarchy
\cite{Barry:2011wb,1110.5795},
since in that case their contribution to the effective Majorana mass is
\cite{1203.5250}
\begin{equation}
1.4 \times 10^{-2}
\lesssim
m_{\beta\beta}^{(\text{light})}
\lesssim
5.0 \times 10^{-2}
\,
\text{eV}
\qquad
(\text{IH} -95\%\,\text{CL})
\,.
\label{mbb-ih}
\end{equation}
On the other hand, no cancellation is possible in the case of a normal hierarchy,
for which
\cite{1203.5250}
\begin{equation}
m_{\beta\beta}^{(\text{light})}
\lesssim
4.5 \times 10^{-3}
\,
\text{eV}
\qquad
(\text{NH} -95\%\,\text{CL})
\,,
\label{mbb-nh}
\end{equation}
and the contribution of $m_{\beta\beta}^{(4)}$ is dominant if it is larger than about $10^{-2} \, \text{eV}$.
Figure~\ref{fig:mbb}
shows the allowed values of $m_{\beta\beta}$
as functions of the lightest neutrino mass
in the two 3+1 schemes
with a normal (left)
and
inverted (right)
mass spectra of the three light neutrinos.
We have drawn also the curves which delimit the three-neutrino mixing
allowed regions
\cite{1203.5250}.
One can see that practically the situation is reversed with respect to the three-neutrino mixing case
(see also the discussion in \cite{1206.2560}),
in which $m_{\beta\beta}$ is predicted to be large in the inverted spectrum
and can vanish in the normal spectrum.
In the 3+1 case $m_{\beta\beta}$ can have any value in the inverted spectrum,
whereas in the normal spectrum it is likely to be large if the three light neutrino masses are hierarchical,
i.e. if $m_{1} \ll m_{2} \ll m_{3}$.

\begin{figure}[t]
\begin{center}
\includegraphics*[width=\linewidth]{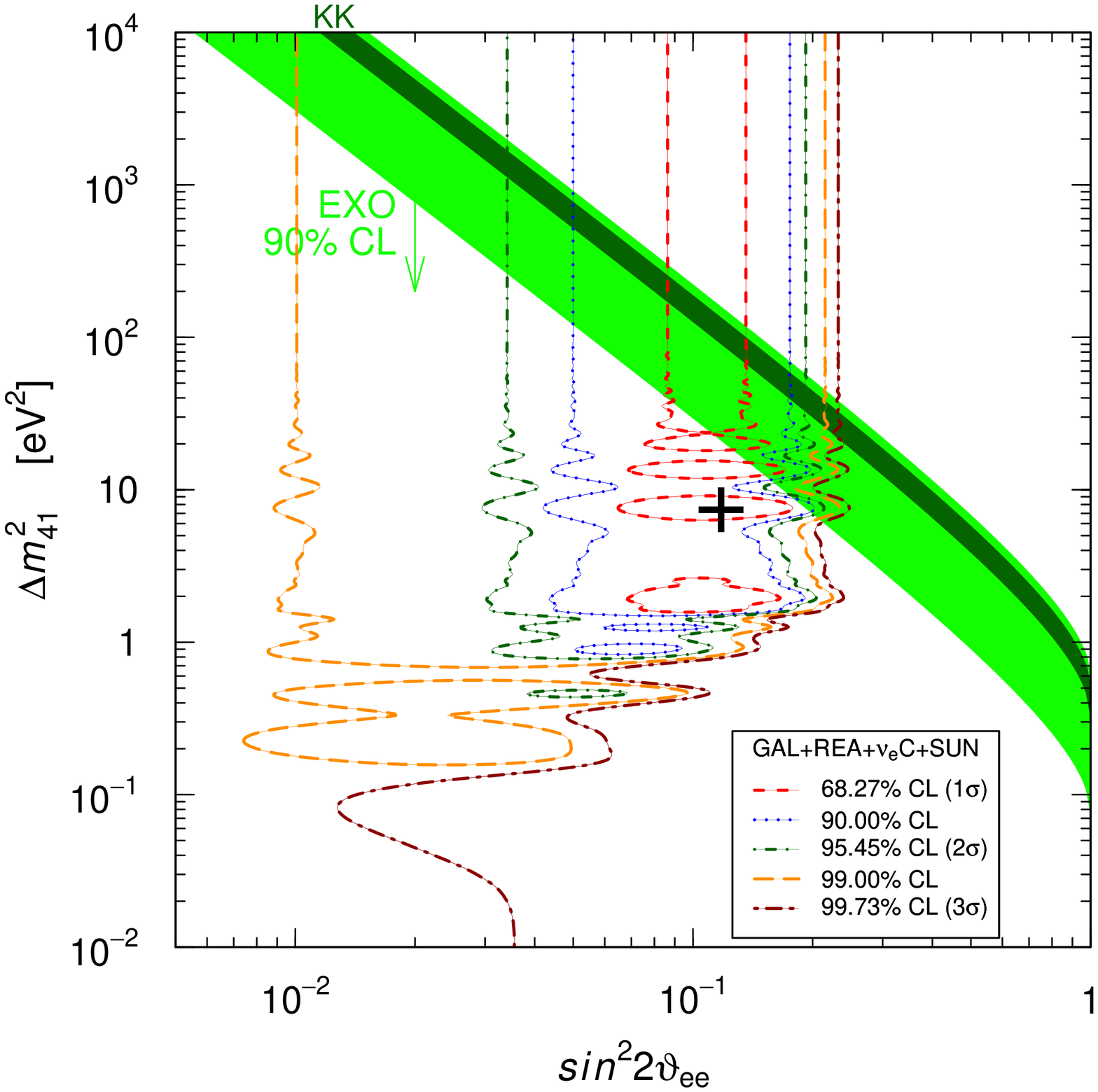}
\end{center}
\caption{ \label{fig:fit-mbb}
Comparison of the
allowed regions in the
$\sin^{2}2\vartheta_{ee}$--$\Delta{m}^{2}_{41}$ plane
obtained from the global fit of $\nu_{e}$ disappearance data
(same as in Fig.~\ref{fig:fit-glo})
and the bound
in the no-cancellation case (\ref{no-cancellation})
corresponding to the
90\% CL upper bound for $m_{\beta\beta}$
obtained in the EXO experiment taking into account nuclear matrix element uncertainties
\cite{Auger:2012ar}
(green band).
The dark-green band corresponds to the $1\sigma$
Klapdor-Kleingrothaus et al. range of $m_{\beta\beta}$
\cite{KlapdorKleingrothaus:2006ff}.
}
\end{figure}

Let us consider the ``no-cancellation'' case,
in which
\begin{equation}
m_{\beta\beta}
\ge
m_{\beta\beta}^{(4)}
\quad
\text{(no-cancellation)}
\,.
\label{no-cancellation}
\end{equation}
In this case,
as shown in Fig.~\ref{fig:mbb-chi},
large values of $m_{\beta\beta}^{(4)}$
are excluded by the currently most stringent upper bound for $m_{\beta\beta}$
obtained in the EXO experiment
\cite{Auger:2012ar}
(the vertical green band in Fig.~\ref{fig:mbb-chi} is the 90\% CL EXO bound
taking into account nuclear matrix element uncertainties).
This limit implies the upper bound on
$\Delta{m}^2_{41} \simeq ( m_{\beta\beta}^{(4)} / |U_{e4}|^2)^2$
as a function of
$\sin^{2}2\vartheta_{ee}$
shown in Fig.~\ref{fig:fit-mbb}.
One can see that parts of the high-$\Delta{m}^2_{41}$ regions allowed by the global fit
are disfavored by the EXO bound.
However,
the large nuclear matrix element uncertainties do not allow to establish a precise bound.
From Fig.~\ref{fig:fit-mbb} one can also see that the putative
Klapdor-Kleingrothaus et al. $1\sigma$ range of $m_{\beta\beta}$
\cite{KlapdorKleingrothaus:2006ff}
implies a rather large value of
$\Delta{m}^2_{41}$, around
$100 - 200 \, \text{eV}^2$.
In this case the oscillation length is very short,
of the order of 1 cm for neutrinos with energy of the order of 1 MeV,
as reactor neutrinos and neutrinos emitted by radioactive sources.
Hence it will be practically impossible to observe a variation of the event rate characteristic of oscillations
in future very short-baseline
reactor neutrino  experiments
\cite{1205.2955,1204.2449}
and 
radioactive source experiments
\cite{hep-ex/9901012,1006.2103,1107.2335,1205.4419}
(see also \cite{Abazajian:2012ys,1209.5090}).

Finally,
considering cosmological measurements one must say that they are a powerful probe of
the number of neutrinos
and of neutrino masses at the eV scale
(see \cite{Wong:2011ip,Abazajian:2012ys,Steigman:2012ve}),
but the analysis requires many assumptions on the cosmological model and its details.
A comparison of the results of the fit of short-baseline oscillation data is beyond the scope of this paper.
We can only say that
the analysis of cosmological data
in the framework of the standard $\Lambda$CDM
\cite{0910.0008,1006.3795,1006.5276,1108.4136,1109.2767,1207.6515,1208.4354}
allow the existence of a sterile neutrino thermalized in the early Universe,
but restricts its mass to be less than about 1 eV
(a possible suppression of the sterile neutrino thermalization
with a large lepton asymmetry has been discussed recently in
\cite{1204.5861,1206.1046}).
If this constraint is correct,
the upper bound on $\Delta{m}^{2}_{41}$ is about $1\,\text{eV}^2$,
which is much more restrictive than those of
$\beta$ decay in Fig.~\ref{fig:con-mnz}
and
neutrinoless double-$\beta$ decay in Fig.~\ref{fig:fit-mbb}.

\section{Conclusions}
\label{Conclusions}

In this paper we presented a complete update of the analysis of
$\nu_{e}$ and $\bar\nu_{e}$ disappearance experiments
in terms of neutrino oscillations
in the framework of 3+1 neutrino mixing.
We have shown that
the Gallium anomaly,
the reactor anomaly,
solar neutrino data
and
$\nu_{e}\text{C}$ scattering data
are compatible with
short-baseline oscillations
with an amplitude
$\sin^{2}2\vartheta_{ee}$
between about 0.03 and 0.2
and a squared-mass difference $\Delta{m}^2_{41}$ larger than about
$0.5 \, \text{eV}^2$
at 95\% CL.
Assuming the mass hierarchy in Eq.~(\ref{mass-hierarchy}),
we have shown that
the heavy neutrino mass $m_{4}$ is observable in
$\beta$-decay experiments
and
neutrinoless double-$\beta$ decay experiments.
The very recent
Mainz $\beta$-decay data
\cite{1210.4194}
constrain $\Delta{m}^2_{41}$ to be smaller than about $10^{4}\,\text{eV}^2$
at 95\% CL.
For Majorana neutrinos,
the recent EXO limit on the effective Majorana mass in
neutrinoless double-$\beta$ decay
\cite{Auger:2012ar}
give a more stringent constraint which can vary between about $10^2$ and $10^3\,\text{eV}^2$
depending on the nuclear matrix element uncertainties
if there are no cancellations between the contribution of $\nu_{4}$
and that of the three light neutrinos.
We think that our results are interesting
for the many projects which will search in the next years effects of light sterile neutrinos
with
electron neutrino and antineutrino radioactive sources
(see \cite{Abazajian:2012ys,Ianni-NOW2012,Link-NOW2012}),
reactor electron antineutrinos
(see \cite{Abazajian:2012ys,Gaffiot-NOW2012})
and
accelerator electron neutrinos
\cite{Antonello:2012hf,Stanco-NOW2012,Gibin-NOW2012}.

\bigskip
\centerline{\textbf{Acknowledgments}}
\medskip

The work of Y.F. Li is supported in part by the National Natural Science Foundation
of China under grant No.~11135009.
The work of H.W. Long is supported in part by the National Natural Science Foundation
of China under grant No.~11265006.


\begin{thebibliography}{100}

\bibitem{Giunti:2007ry}
C.~Giunti and C.~W. Kim,
{\em {Fundamentals of Neutrino Physics and Astrophysics}} (Oxford
University Press, Oxford, UK, 2007),
{ISBN 978-0-19-850871-7}.

\bibitem{Bilenky:2010zza}
S.~Bilenky,
{\em {Introduction to the physics of massive and mixed neutrinos}}
(Springer, 2010),
{Lecture Notes in Physics, Volume 817; ISBN 978-3-642-14042-6}.

\bibitem{Xing:2011zza}
Z.-z. Xing and S.~Zhou,
{\em {Neutrinos in particle physics, astronomy and cosmology}}
(Zhejiang University Press, 2011),
{ISBN 978-7-308-08024-8}.

\bibitem{Laveder:2007zz}
M.~Laveder,
Nucl. Phys. Proc. Suppl. {\bf 168}, 344 (2007),
{Workshop on Neutrino Oscillation Physics (NOW 2006), Otranto, Lecce,
Italy, 9-16 Sep 2006}.

\bibitem{hep-ph/0610352}
C.~Giunti and M.~Laveder,
Mod. Phys. Lett. {\bf A22}, 2499 (2007), hep-ph/0610352.

\bibitem{Giunti:2010zu}
C.~Giunti and M.~Laveder,
Phys. Rev. {\bf C83}, 065504 (2011), arXiv:1006.3244.

\bibitem{Mueller:2011nm}
T.~A. Mueller {\em et~al.},
Phys. Rev. {\bf C83}, 054615 (2011), arXiv:1101.2663.

\bibitem{Huber:2011wv}
P.~Huber,
Phys. Rev. {\bf C84}, 024617 (2011), arXiv:1106.0687.

\bibitem{Mention:2011rk}
G.~Mention {\em et~al.},
Phys. Rev. {\bf D83}, 073006 (2011), arXiv:1101.2755.

\bibitem{hep-ex/0104049}
LSND, A.~Aguilar {\em et~al.},
Phys. Rev. {\bf D64}, 112007 (2001), hep-ex/0104049.

\bibitem{1207.4809}
MiniBooNE, A.~A. Aguilar-Arevalo {\em et~al.},
(2012), arXiv:1207.4809.

\bibitem{Giunti-Laveder-Li-Long-12}
C.~Giunti, M.~Laveder, Y.~Li, Q.~Liu, and H.~Long,
(2012),
{In Preparation}.

\bibitem{hep-ex/0509008}
ALEPH, DELPHI, L3, OPAL, SLD, LEP Electroweak Working Group, SLD Electroweak
Group, SLD Heavy Flavour Group, S.~Schael {\em et~al.},
Phys. Rept. {\bf 427}, 257 (2006), hep-ex/0509008.

\bibitem{Frekers:2011zz}
D.~Frekers {\em et~al.},
Phys. Lett. {\bf B706}, 134 (2011).

\bibitem{Abazajian:2012ys}
K.~N. Abazajian {\em et~al.},
(2012), arXiv:1204.5379.

\bibitem{1210.4194}
C.~Kraus, A.~Singer, K.~Valerius, and C.~Weinheimer,
(2012), arXiv:1210.4194.

\bibitem{Auger:2012ar}
EXO Collaboration, M.~Auger {\em et~al.},
(2012), arXiv:1205.5608.

\bibitem{KlapdorKleingrothaus:2006ff}
H.~V. Klapdor-Kleingrothaus and I.~V. Krivosheina,
Mod. Phys. Lett. {\bf A21}, 1547 (2006).

\bibitem{hep-ph/9812360}
S.~M. Bilenky, C.~Giunti, and W.~Grimus,
Prog. Part. Nucl. Phys. {\bf 43}, 1 (1999), hep-ph/9812360.

\bibitem{hep-ph/0405172}
M.~Maltoni, T.~Schwetz, M.~Tortola, and J.~Valle,
New J. Phys. {\bf 6}, 122 (2004), hep-ph/0405172.

\bibitem{hep-ph/0606054}
A.~Strumia and F.~Vissani,
(2006), hep-ph/0606054.

\bibitem{GonzalezGarcia:2007ib}
M.~C. Gonzalez-Garcia and M.~Maltoni,
Phys. Rept. {\bf 460}, 1 (2008), arXiv:0704.1800.

\bibitem{Giunti:2009xz}
C.~Giunti and Y.~Li,
Phys. Rev. {\bf D80}, 113007 (2009), arXiv:0910.5856.

\bibitem{Palazzo:2011rj}
A.~Palazzo,
Phys. Rev. {\bf D83}, 113013 (2011), arXiv:1105.1705.

\bibitem{Palazzo:2012yf}
A.~Palazzo,
Phys. Rev. {\bf D85}, 077301 (2012), arXiv:1201.4280.

\bibitem{Palazzo-NOW2012}
A.~Palazzo,
(2012),
{NOW 2012, Neutrino Oscillation Workshop, 9-16 September 2012, Conca
Specchiulla, Otranto, Italy}.

\bibitem{1106.5552}
J.~Conrad and M.~Shaevitz,
Phys. Rev. {\bf D85}, 013017 (2012), arXiv:1106.5552.

\bibitem{Giunti:2011cp}
C.~Giunti and M.~Laveder,
Phys. Lett. {\bf B706}, 200 (2011), arXiv:1111.1069.

\bibitem{Galeazzi:2001py}
M.~Galeazzi, F.~Fontanelli, F.~Gatti, and S.~Vitale,
Phys. Rev. Lett. {\bf 86}, 1978 (2001).

\bibitem{Anselmann:1995ar}
GALLEX, P.~Anselmann {\em et~al.},
Phys. Lett. {\bf B342}, 440 (1995).

\bibitem{Hampel:1998fc}
GALLEX, W.~Hampel {\em et~al.},
Phys. Lett. {\bf B420}, 114 (1998).

\bibitem{1001.2731}
F.~Kaether, W.~Hampel, G.~Heusser, J.~Kiko, and T.~Kirsten,
Phys. Lett. {\bf B685}, 47 (2010), arXiv:1001.2731.

\bibitem{Abdurashitov:1996dp}
SAGE, J.~N. Abdurashitov {\em et~al.},
Phys. Rev. Lett. {\bf 77}, 4708 (1996).

\bibitem{hep-ph/9803418}
SAGE, J.~N. Abdurashitov {\em et~al.},
Phys. Rev. {\bf C59}, 2246 (1999), hep-ph/9803418.

\bibitem{nucl-ex/0512041}
J.~N. Abdurashitov {\em et~al.},
Phys. Rev. {\bf C73}, 045805 (2006), nucl-ex/0512041.

\bibitem{0901.2200}
SAGE, J.~N. Abdurashitov {\em et~al.},
Phys. Rev. {\bf C80}, 015807 (2009), arXiv:0901.2200.

\bibitem{Bahcall:1997eg}
J.~N. Bahcall,
Phys. Rev. {\bf C56}, 3391 (1997), hep-ph/9710491.

\bibitem{Krofcheck:1985fg}
D.~Krofcheck {\em et~al.},
Phys. Rev. Lett. {\bf 55}, 1051 (1985).

\bibitem{Krofcheck-PhD-1987}
D.~{Krofcheck},
(1987),
{PhD Thesis, Ohio State University}.

\bibitem{Haxton:1998uc}
W.~C. Haxton,
Phys. Lett. {\bf B431}, 110 (1998), nucl-th/9804011.

\bibitem{Bahcall:1989ks}
J.~N. Bahcall,
{\em {Neutrino Astrophysics}} (Cambridge University Press, 1989).

\bibitem{PDG-2012}
Particle Data Group, J.~Beringer {\em et~al.},
Phys. Rev. D {\bf 86}, 010001 (2012).

\bibitem{Hampel:1985zz}
W.~Hampel and L.~Remsberg,
Phys.Rev. {\bf C31}, 666 (1985).

\bibitem{LOGFT-2012}
{National Nuclear Data Center, Brookhaven National Laboratory},
(2012),
{\url{http://www.nndc.bnl.gov/logft/}}.

\bibitem{1008.4750}
C.~Giunti and M.~Laveder,
Phys. Rev. {\bf D82}, 113009 (2010), arXiv:1008.4750.

\bibitem{1107.1452}
C.~Giunti and M.~Laveder,
Phys.Rev. {\bf D84}, 073008 (2011), arXiv:1107.1452.

\bibitem{1109.4033}
C.~Giunti and M.~Laveder,
Phys.Rev. {\bf D84}, 093006 (2011), arXiv:1109.4033.

\bibitem{1207.6515}
M.~Archidiacono, N.~Fornengo, C.~Giunti, and A.~Melchiorri,
Phys. Rev. {\bf D86}, 065028 (2012), arXiv:1207.6515.

\bibitem{Declais:1995su}
Bugey, B.~Achkar {\em et~al.},
Nucl. Phys. {\bf B434}, 503 (1995).

\bibitem{Declais:1994ma}
Bugey, Y.~Declais {\em et~al.},
Phys. Lett. {\bf B338}, 383 (1994).

\bibitem{Kuvshinnikov:1990ry}
A.~Kuvshinnikov, L.~Mikaelyan, S.~Nikolaev, M.~Skorokhvatov, and A.~Etenko,
JETP Lett. {\bf 54}, 253 (1991).

\bibitem{Zacek:1986cu}
CalTech-SIN-TUM, G.~Zacek {\em et~al.},
Phys. Rev. {\bf D34}, 2621 (1986).

\bibitem{Hoummada:1995zz}
A.~Hoummada, S.~Lazrak~Mikou, G.~Bagieu, J.~Cavaignac, and D.~Holm~Koang,
Applied Radiation and Isotopes {\bf 46}, 449 (1995).

\bibitem{Vidyakin:1990iz}
Krasnoyarsk, G.~S. Vidyakin {\em et~al.},
Sov. Phys. JETP {\bf 71}, 424 (1990).

\bibitem{hep-ph/0304176}
M.~Maltoni and T.~Schwetz,
Phys. Rev. {\bf D68}, 033020 (2003), hep-ph/0304176.

\bibitem{Cleveland:1998nv}
Homestake, B.~T. Cleveland {\em et~al.},
Astrophys. J. {\bf 496}, 505 (1998).

\bibitem{astro-ph/0204245}
SAGE, J.~N. Abdurashitov {\em et~al.},
J. Exp. Theor. Phys. {\bf 95}, 181 (2002), astro-ph/0204245.

\bibitem{hep-ex/0508053}
Super-Kamkiokande, J.~Hosaka {\em et~al.},
Phys. Rev. {\bf D73}, 112001 (2006), hep-ex/0508053.

\bibitem{0803.4312}
Super-Kamiokande, J.~Cravens {\em et~al.},
Phys. Rev. {\bf D78}, 032002 (2008), arXiv:0803.4312.

\bibitem{1010.0118}
Super-Kamiokande, K.~Abe {\em et~al.},
Phys. Rev. {\bf D83}, 052010 (2011), arXiv:1010.0118.

\bibitem{Smy-NU2012}
Super-Kamiokande, M.~Smy,
(2012),
{Neutrino 2012, XXV International Conference on Neutrino Physics and
Astrophysics, 3-9 June 2012, Kyoto, Japan}.

\bibitem{nucl-ex/0204009}
SNO, Q.~R. Ahmad {\em et~al.},
Phys. Rev. Lett. {\bf 89}, 011302 (2002), nucl-ex/0204009.

\bibitem{nucl-ex/0502021}
SNO, B.~Aharmim {\em et~al.},
Phys. Rev. {\bf C72}, 055502 (2005), nucl-ex/0502021.

\bibitem{0806.0989}
SNO, B.~Aharmim {\em et~al.},
Phys. Rev. Lett. {\bf 101}, 111301 (2008), arXiv:0806.0989.

\bibitem{1104.1816}
Borexino, G.~Bellini {\em et~al.},
Phys. Rev. Lett. {\bf 107}, 141302 (2011), arXiv:1104.1816.

\bibitem{1110.3230}
Borexino, G.~Bellini {\em et~al.},
Phys. Rev. Lett. {\bf 108}, 051302 (2012), arXiv:1110.3230.

\bibitem{1009.4771}
KamLAND, A.~Gando {\em et~al.},
Phys. Rev. {\bf D83}, 052002 (2011), arXiv:1009.4771.

\bibitem{Dooling:1999sg}
D.~Dooling, C.~Giunti, K.~Kang, and C.~W. Kim,
Phys. Rev. {\bf D61}, 073011 (2000), hep-ph/9908513.

\bibitem{Giunti:2000wt}
C.~Giunti, M.~C. Gonzalez-Garcia, and C.~Pena-Garay,
Phys. Rev. {\bf D62}, 013005 (2000), hep-ph/0001101.

\bibitem{1203.1669}
Daya Bay, F.~P. An {\em et~al.},
Phys. Rev. Lett. {\bf 108}, 171803 (2012), arXiv:1203.1669.

\bibitem{1204.0626}
RENO, S.-B. Kim {\em et~al.},
Phys. Rev. Lett. {\bf 108}, 191802 (2012), arXiv:1204.0626.

\bibitem{Bilenky:1996rw}
S.~M. Bilenky, C.~Giunti, and W.~Grimus,
Eur. Phys. J. {\bf C1}, 247 (1998), hep-ph/9607372.

\bibitem{0809.5076}
A.~de~Gouvea and T.~Wytock,
Phys. Rev. {\bf D79}, 073005 (2009), arXiv:0809.5076.

\bibitem{Giunti:2011vc}
C.~Giunti and M.~Laveder,
Phys. Rev. {\bf D85}, 031301 (2012), arXiv:1111.5211.

\bibitem{0910.2984}
SNO, B.~Aharmim {\em et~al.},
Phys. Rev. {\bf C81}, 055504 (2010), arXiv:0910.2984.

\bibitem{1109.0763}
SNO, B.~Aharmim {\em et~al.},
(2011), arXiv:1109.0763.

\bibitem{Bahcall:2004fg}
J.~N. Bahcall and M.~H. Pinsonneault,
Phys. Rev. Lett. {\bf 92}, 121301 (2004), astro-ph/0402114.

\bibitem{Bodmann:1994py}
KARMEN., B.~E. Bodmann {\em et~al.},
Phys. Lett. {\bf B332}, 251 (1994).

\bibitem{hep-ex/9801007}
KARMEN, B.~Armbruster {\em et~al.},
Phys. Rev. {\bf C57}, 3414 (1998), hep-ex/9801007.

\bibitem{hep-ex/0105068}
LSND, L.~B. Auerbach {\em et~al.},
Phys. Rev. {\bf C64}, 065501 (2001), hep-ex/0105068.

\bibitem{Otten:2008zz}
E.~Otten and C.~Weinheimer,
Rept.Prog.Phys. {\bf 71}, 086201 (2008), arXiv:0909.2104.

\bibitem{Wolf-NPB2012}
KATRIN, J.~Wolf,
(2012),
{NPB 2012, International Symposium on Neutrino Physics and Beyond,
23-26 September 2012, Shenzhen, China}.

\bibitem{Esmaili:2012vg}
A.~Esmaili and O.~L.~G. Peres,
Phys. Rev. {\bf D85}, 117301 (2012), arXiv:1203.2632.

\bibitem{1109.5515}
J.~Gomez-Cadenas, J.~Martin-Albo, M.~Mezzetto, F.~Monrabal, and M.~Sorel,
Riv. Nuovo Cim. {\bf 35}, 29 (2012), arXiv:1109.5515.

\bibitem{1203.5250}
S.~M. Bilenky and C.~Giunti,
Mod. Phys. Lett. {\bf A27}, 1230015 (2012), arXiv:1203.5250.

\bibitem{1205.0649}
J.~D. Vergados, H.~Ejiri, and F.~Simkovic,
Rept. Prog. Phys. {\bf 75}, 106301 (2012), arXiv:1205.0649.

\bibitem{1206.2560}
W.~Rodejohann,
J. Phys. {\bf G39}, 124008 (2012), arXiv:1206.2560.

\bibitem{Cremonesi-NOW2012}
O.~Cremonesi,
(2012),
{NOW 2012, Neutrino Oscillation Workshop, 9-16 September 2012, Conca
Specchiulla, Otranto, Italy}.

\bibitem{Barry:2011wb}
J.~Barry, W.~Rodejohann, and H.~Zhang,
JHEP {\bf 07}, 091 (2011), arXiv:1105.3911.

\bibitem{1110.5795}
Y.~Li and S.~Liu,
Phys. Lett. {\bf B706}, 406 (2012), arXiv:1110.5795.

\bibitem{1205.2955}
A.~P. Serebrov {\em et~al.},
(2012), arXiv:1205.2955.

\bibitem{1204.2449}
A.~V. Derbin, A.~S. Kayunov, and V.~N. Muratova,
(2012), arXiv:1204.2449.

\bibitem{hep-ex/9901012}
A.~Ianni, D.~Montanino, and G.~Scioscia,
Eur. Phys. J. {\bf C8}, 609 (1999), hep-ex/9901012.

\bibitem{1006.2103}
V.~N. Gavrin, V.~V. Gorbachev, E.~P. Veretenkin, and B.~T. Cleveland,
(2010), arXiv:1006.2103.

\bibitem{1107.2335}
M.~Cribier {\em et~al.},
Phys. Rev. Lett. {\bf 107}, 201801 (2011), arXiv:1107.2335.

\bibitem{1205.4419}
A.~Bungau {\em et~al.},
(2012), arXiv:1205.4419.

\bibitem{1209.5090}
T.~Lasserre,
(2012), arXiv:1209.5090,
{Neutrino 2012 Conference, Kyoto, Japan, June 2012}.

\bibitem{Wong:2011ip}
Y.~Y.~Y. Wong,
Ann. Rev. Nucl. Part. Sci. {\bf 61}, 69 (2011), arXiv:1111.1436.

\bibitem{Steigman:2012ve}
G.~Steigman,
Adv. High Energy Phys. {\bf 2012}, 268321 (2012), arXiv:1208.0032.

\bibitem{0910.0008}
B.~A. Reid, L.~Verde, R.~Jimenez, and O.~Mena,
JCAP {\bf 1001}, 003 (2010), arXiv:0910.0008.

\bibitem{1006.3795}
M.~C. Gonzalez-Garcia, M.~Maltoni, and J.~Salvado,
JHEP {\bf 08}, 117 (2010), arXiv:1006.3795.

\bibitem{1006.5276}
J.~Hamann, S.~Hannestad, G.~G. Raffelt, I.~Tamborra, and Y.~Y. Wong,
Phys. Rev. Lett. {\bf 105}, 181301 (2010), arXiv:1006.5276.

\bibitem{1108.4136}
J.~Hamann, S.~Hannestad, G.~G. Raffelt, and Y.~Y. Wong,
JCAP {\bf 1109}, 034 (2011), arXiv:1108.4136.

\bibitem{1109.2767}
M.~Archidiacono, E.~Calabrese, and A.~Melchiorri,
Phys. Rev. {\bf D84}, 123008 (2011), arXiv:1109.2767.

\bibitem{1208.4354}
S.~Joudaki, K.~N. Abazajian, and M.~Kaplinghat,
Phys. Rev. {\bf D87}, 065003 (2013), arXiv:1208.4354.

\bibitem{1204.5861}
S.~Hannestad, I.~Tamborra, and T.~Tram,
JCAP {\bf 1207}, 025 (2012), arXiv:1204.5861.

\bibitem{1206.1046}
A.~Mirizzi, N.~Saviano, G.~Miele, and P.~D. Serpico,
Phys.Rev. {\bf D86}, 053009 (2012), arXiv:1206.1046.

\bibitem{Ianni-NOW2012}
A.~Ianni,
(2012),
{NOW 2012, Neutrino Oscillation Workshop, 9-16 September 2012, Conca
Specchiulla, Otranto, Italy}.

\bibitem{Link-NOW2012}
J.~Link,
(2012),
{NOW 2012, Neutrino Oscillation Workshop, 9-16 September 2012, Conca
Specchiulla, Otranto, Italy}.

\bibitem{Gaffiot-NOW2012}
J.~Gaffiot,
(2012),
{NOW 2012, Neutrino Oscillation Workshop, 9-16 September 2012, Conca
Specchiulla, Otranto, Italy}.

\bibitem{Antonello:2012hf}
M.~Antonello {\em et~al.},
(2012), arXiv:1203.3432.

\bibitem{Stanco-NOW2012}
L.~Stanco,
(2012),
{NOW 2012, Neutrino Oscillation Workshop, 9-16 September 2012, Conca
Specchiulla, Otranto, Italy}.

\bibitem{Gibin-NOW2012}
D.~Gibin,
(2012),
{NOW 2012, Neutrino Oscillation Workshop, 9-16 September 2012, Conca
Specchiulla, Otranto, Italy}.

\end{thebibliography}

\end{document}